\documentclass{llncs}

\usepackage{chngpage}

\usepackage{booktabs}

\usepackage[utf8]{inputenc}

\usepackage{url}
\usepackage{amsmath, amssymb}
\usepackage{examplep}
\usepackage{mathabx}

\newcommand{\ms}[1]{\texttt{#1}}

\usepackage{fancyvrb}
\DefineVerbatimEnvironment{ex}{Verbatim}{numbers=left,numbersep=2mm,frame=single,fontsize=\scriptsize}
\DefineVerbatimEnvironment{dl}{}{fontsize=\scriptsize}

\newenvironment{gcotable-old}{
  \scriptsize
  \sffamily
  \vspace{0cm}\linebreak
  \begin{tabular}{l|l|l|l|l|l}
  \hline
  \textbf{c. type} & \textbf{context concept} & \textbf{p. list} & \textbf{concepts} & \textbf{c. element} & \textbf{c. value} \\
  \hline

}{
  \hline
  \end{tabular}
  \linebreak
}

\newenvironment{gcotable}{
  \scriptsize
  \sffamily
  \vspace{0cm}
	\begin{center}
	\textbf{\vspace{0.4cm}Mapping to RDF-CO:} \\
  \begin{tabular}{l|l|l|l|l|l|l}
	\hline
  \textbf{c. set} & \textbf{context class} & \textbf{left p. list} & \textbf{right p. list} & \textbf{classes} & \textbf{c. element} & \textbf{c. value} \\
  \hline

}{
  \hline
  \end{tabular}
	\end{center}
}

\newenvironment{DL}{
  \vspace{0cm}
	\begin{center}
  \begin{tabular}{r l}

}{
  \end{tabular}
	\end{center}
}

\newenvironment{evaluation-overall}{
  \scriptsize
  \sffamily
  \vspace{0cm}
	\begin{center}
  \begin{tabular}{l|c|c|c|c|c|c}
  \hline
  \textbf{constraint} & \textbf{DSP} & \textbf{OWL2-DL} & \textbf{OWL2-QL} & \textbf{ReSh} & \textbf{ShEx} & \textbf{SPIN} \\
  \hline

}{
  \hline
  \end{tabular}
  \linebreak
	\end{center}
}

\newenvironment{evaluation-generic}{
  \scriptsize
  \sffamily
  \vspace{0cm}
	\begin{center}
  \begin{tabular}{l|c|c|c}
  \hline
  \textbf{constraint} & \textbf{property c.} & \textbf{simple c.} & \textbf{DL} \\
  \hline

}{
  \hline
  \end{tabular}
  \linebreak
	\end{center}
}

\newenvironment{evaluation-generic-overview}{
  \scriptsize
  \sffamily
  \vspace{0cm}
	\begin{center}
  \begin{tabular}{l|c|c}
  \hline
  \textbf{Constraint Classes} & \textbf{\#} & \textbf{\%} \\
  \hline

}{
  \hline
  \end{tabular}
  \linebreak
	\end{center}
}

\newenvironment{evaluation-cwa-una-dependency}{
  \scriptsize
  \sffamily
  \vspace{0cm}
	\begin{center}
  \begin{tabular}{l|c}
  \hline
  \textbf{Constraint Type} & \textbf{CWA/UNA Dependent} \\
  \hline

}{
  \hline
  \end{tabular}
  \linebreak
	\end{center}
}

\newenvironment{cwa-una-dependency}{
  \scriptsize
  \sffamily
  \vspace{0cm}
	\begin{center}
  \begin{tabular}{l|c|c}
  \hline
  \textbf{Constraint Type} & \textbf{CWA/UNA Dependent} & \textbf{CWA/UNA Dependent} \\
  \hline

}{
  \hline
  \end{tabular}
  \linebreak
	\end{center}
}

\usepackage{xspace}

\usepackage{tikz}
\def\checkmark{\tikz\fill[scale=0.4](0,.35) -- (.25,0) -- (1,.7) -- (.25,.15) -- cycle;} 

\usepackage{pifont}

\usepackage{tabularx}

\usepackage[colorinlistoftodos]{todonotes}

\usepackage{array,graphicx}
\usepackage{booktabs}
\usepackage{pifont}
\newcommand*\rot{\rotatebox{90}}

\usepackage{booktabs}

\usepackage{tablefootnote}

\usepackage{float}

\begin{document}

\title{RDF Validation Requirements - Evaluation and Logical Underpinning}
\subtitle{}
\titlerunning{XXXXX}  
%
\author{Thomas Bosch\inst{1} \and Andreas Nolle\inst{2} \and Erman Acar\inst{3} \and Kai Eckert \inst{3}}
\authorrunning{XXXXX} 
%
\institute{GESIS – Leibniz Institute for the Social Sciences, Germany\\
\email{thomas.bosch@gesis.org},\\ 
\and
Albstadt-Sigmaringen University, Germany \\
\email{nolle@hs-albsig.de}
\and
University of Mannheim, Germany \\
\email{\{erman, kai\}@informatik.uni-mannheim.de}}

\maketitle              

\begin{abstract}
There are many case studies for which the formulation of RDF constraints and the validation of RDF data conforming to these constraint is very important. As a part of the collaboration with the W3C and the DCMI working groups on RDF validation, we identified major RDF validation requirements
and initiated an RDF validation requirements database which is available to contribute at \url{http://purl.org/net/rdf-validation}.
The purpose of this database is to collaboratively collect case studies, use cases, requirements, and solutions regarding RDF validation. Although, there are multiple constraint languages which can be used to formulate RDF constraints (associated with these requirements), 
there is no standard way to formulate them. This paper serves to evaluate to which extend each requirement is satisfied by each of these constraint languages. We take reasoning into account  as an important pre-validation step and therefore map constraints to DL
in order to show that each constraint can be mapped to an ontology describing RDF constraints generically.

\end{abstract}
%


\section{Introduction}

The W3C organized the RDF Validation Workshop\footnote{\url{http://www.w3.org/2012/12/rdf-val/}}, 
where experts from industry, government, and academia discussed first use cases for RDF constraint formulation and validation.
In 2014, two working groups on RDF validation have been established: 
the W3C RDF Data Shapes\footnote{\url{http://www.w3.org/2014/rds/charter}} and the DCMI RDF Application Profiles working groups\footnote{\url{http://wiki.dublincore.org/index.php/RDF-Application-Profiles}}. 
Bosch and Eckert \cite{BoschEckert2014} collected the findings of these working groups and initiated a database of requirements to formulate and validate RDF constraints.
The database is available for contribution at \url{http://purl.org/net/rdf-validation}.
The intention associated with this database is to collaboratively collect case studies, use cases, requirements, and solutions regarding RDF validation in a comprehensive and structured way. 
The requirements are classified to better evaluate existing solutions. 
We mapped each requirement to formulate RDF constraints directly to an RDF constraint type.

A \ms{constraint language} is a language which is used to formulate constraints.
The W3C Data Shapes working group defines \ms{constraint} as a component of a schema what needs to be satisfied\footnote{\url{https://www.w3.org/2014/data-shapes/wiki/Glossary}}.
There is no constraint language which can be seen as the standalone standard.
However, there are multiple constraint languages (each having its own  syntax and semantics) which can be used to express RDF constraints, such as existential and universal quantification, cardinality restrictions, and exclusive-or of properties.
The five most popular constraint languages are 
Description Set Profiles (DSP)\footnote{\url{http://dublincore.org/documents/2008/03/31/dc-dsp/}},
Resource Shapes (ReSh)\footnote{\url{http://www.w3.org/Submission/shapes/}}, 
Shape Expressions (ShEx)\footnote{\url{http://www.w3.org/Submission/shex-primer/}},
the SPARQL Inferencing Notation (SPIN)\footnote{\url{http://spinrdf.org/}}, 
and the Web Ontology Language (OWL 2)\footnote{\url{http://www.w3.org/TR/owl2-syntax/}}.

In this paper, we describe each requirement within the RDF validation requirements database in detail (sections 2-75).
Additional descriptions can be found directly in the database.
Each requirement corresponds to an RDF constraint type which may be expressible by multiple constraint languages.
For each requirement, we represent some examples in different constraint languages.

We evaluated to which extend the most promising five constraint languages fulfill each of the overall 74 requirements to formulate RDF constraints (section \ref{sec:evaluation-1}).
We distinguished if a constraint is fulfilled by OWL 2 QL or if the more expressive OWL 2 DL is needed. 
We also take reasoning into account, as reasoning may be performed prior to validating constraints.

In order to define an ontology to describe RDF constraints generically, it is needed to define the terminology for the formulation of RDF constraints and to classify them. 
We identified four dimensions to classify constraints:
\begin{itemize}
  \item \ms{Universality:} specific constraints vs. generic constraints
	\item \ms{Complexity:} simple constraints vs. complex constraints
	\item \ms{Context:} property constraints vs. class constraints
	\item \ms{DL Expressivity:} constraints expressible in DL vs. constraints not expressible in DL
\end{itemize}

As there are already five promising constraint languages, our purpose is not to invent a new constraint language.
We rather developed a very simple ontology (only three classes, three object properties, and three data properties) which is universal enough to describe any RDF constraint expressible by any RDF constraint language.
We call this ontology the \ms{RDF constraints ontology (RDF-CO)}\footnote{Available at: \url{https://github.com/boschthomas/RDF-CO}}.

\ms{Specific constraints} are expressed by specific constraint languages like DSP, OWL 2, ReSh, ShEx, and SPIN.
\ms{Generic constraints} are expressed by the RDF-CO.
As RDF-CO describes constraints generically, it does not distinguish constraints according to the dimension \ms{universality}. 
The majority of constraints can be expressed in DL.
In contrast, there are constraints which cannot be expressed in DL, but are also expressible in the RDF-CO. 
\ms{Complex constraints} are built by combining \ms{simple constraints} or complex constraints.
DL statements which represents complex constraints are created out of DL statements representing composed constraints (if expressible in DL). 
Simple constraints may be applied to either properties (\ms{properties constraints}) or classes (\ms{class constraints}).
There are no terms representing simple and complex constraints in the RDF-CO, since context classes (associated simple constraints hold for individuals of these classes) of simple constraints may just be reused by further constraints.
As a consequence, the distinction of property and class constraints is sufficient to describe all possible RDF constraints.

In this paper, we investigate which constraints can be expressed in DL and which not.
If a constraint can be expressed in DL, we added the mapping to DL and to the generic constraint in order to logically underpin associated requirements.
If a constraint cannot be expressed in DL, we only added the mapping to the generic constraint.
Therefore, we show that each constraint can be mapped to a generic constraint.
In section \ref{sec:evaluation-2} we classify the constraints according to the dimensions to classify constraints.

\section{Subsumption}

\ms{Subsumption} (DL terminology: \ms{concept inclusion}) corresponds to the requirement \ms{R-100-SUBSUMPTION}.
A subclass axiom SubClassOf( CE1 CE2 ) states that the class expression CE1 is a subclass of the class expression CE2. 
Roughly speaking, this states that CE1 is more specific than CE2.

\subsection{Simple Example}

All mothers are parents.
The concept \ms{Mother} is subsumed by the concept \ms{Parent}:

\begin{DL}
Mother $\sqsubseteq$ Parent \\
\end{DL}

\begin{gcotable}
class & Mother & - & - & Parent & $\sqsubseteq$ \\
\end{gcotable}

\subsection{Simple Example}

Jedis feel the force:

\begin{DL}
Jedi $\sqsubseteq$ FeelingForce
\end{DL}

Expressed by multiple constraint languages:

\begin{ex}
# OWL2:
Jedi rdfs:subClassOf FeelingForce .
\end{ex}

\begin{ex}
# ReSh:
the extension ext:extendsShape may be used
\end{ex}

\begin{ex}
# ShEx:
FeelingForce {
    feelingForce (true) }
Jedi {
    & FeelingForce ,
    attitute ('good') }
\end{ex}

Data matching the shapes \ms{FeelingForce} and \ms{Jedi}:

\begin{ex}
Yoda 
    feelingForce true ;
    attitute 'good' .
\end{ex}

\begin{gcotable}
class & Jedi & - & - & FeelingForce & $\sqsubseteq$ \\
\end{gcotable}

\subsection{Complex Example}

If an individual is rich, then this individual is not poor:

\begin{DL}
Rich $\sqsubseteq$ $\neg$ Poor 
\end{DL}

\begin{gcotable}
class & $\neg$ Poor & - & - & Poor & $\neg$ \\
class & Rich & - & - & $\neg$ Poor & $\sqsubseteq$ \\
\end{gcotable}

\section{Class Equivalence}

\ms{Class Equivalence} corresponds to the requirement \ms{R-3-EQUIVALENT-CLASSES}.
Concept equivalence asserts that two concepts have the same instances \cite{Kroetzsch2012}.
While synonyms are an obvious example of equivalent concepts, in practice one more
often uses concept equivalence to give a name to complex expressions \cite{Kroetzsch2012}.
Concept equivalence is indeed subsumption from left and right ($A \sqsubseteq B$ and $B \sqsubseteq A$ implies $A \equiv B$ ).

\subsection{Simple Example}

\begin{DL}
Person $\equiv$ Human
\end{DL}

\begin{gcotable}
class & Person & - & - & Human & $\sqsubseteq$ \\
class & Human & - & - & Person & $\sqsubseteq$ \\
\end{gcotable}

\section{Sub Properties}

\ms{Sub Properties} (DL terminology: \ms{role inclusion}) correspond to the requirements \ms{R-54-SUB-OBJECT-PROPERTIES} and \ms{R-54-SUB-DATA-PROPERTIES}.
Subproperty axioms are analogous to subclass axioms.
These axioms state that the property expression PE1 is a subproperty of the property expression PE2 — that is, if an individual x is connected by PE1 to an individual or a literal y, then x is also connected by PE2 to y. 

\subsection{Simple Example}

\begin{DL}
parentOf $\sqsubseteq$ ancestorOf 
\end{DL}

States that \ms{parentOf} is a sub-role of \ms{ancestorOf} , i.e., every pair of individuals related by \ms{parentOf} is also related by \ms{ancestorOf}.

\begin{gcotable}
property & $\top$ & parentOf & ancestorOf & $\top$ & $\sqsubseteq$ \\
\end{gcotable}



\subsection{Simple Example}

\begin{DL}
hasDog $\sqsubseteq$ hasPet 
\end{DL}


\subsection{Complex Example}

\begin{DL}
 $Person \sqcap  \forall hasAge. \leq_{10} \sqsubseteq Person \sqcap \forall hasEvenAge. \leq_{10}$ 
\end{DL}

\section{Object Property Paths}

\ms{Object Property Paths} (or Object Property Chains and in DL terminology complex role inclusion axiom or role composition)
corresponds to the requirement \ms{R-55-OBJECT-PROPERTY-PATHS}.
The more complex form of sub properties. This axiom states that, if an individual x is connected by a sequence of object property expressions OPE1, ..., OPEn with an individual y, then x is also connected with y by the object property expression OPE.  
Role composition can only appear on the left-hand side of complex role inclusions \cite{Kroetzsch2012}.

\subsection{Simple Example}

\begin{DL}
brotherOf $\circ$ parentOf $\sqsubseteq$ uncleOf 
\end{DL}

\begin{ex}
# OWL2:
uncleOf owl:propertyChainAxiom ( brotherOf parentOf ) . 
\end{ex}

\begin{gcotable}
property & $\top$ & brotherOf, parentOf & uncleOf & $\top$ & $\sqsubseteq$ & -  \\
\end{gcotable}

\section{Disjoint Properties}

\ms{Disjoint Properties} corresponds to the requirement \ms{R-9-DISJOINT-PROPERTIES}.
A disjoint properties axiom states that all of the properties are pairwise disjoint; 
that is, no individual x can be connected to an individual y by these properties. 

\subsection{Simple Example}

The object properties \ms{parentOf} and \ms{childOf} are disjoint:

\begin{DL}
Disjoint(parentOf, childOf)\\
\end{DL}

or alternatively:

\begin{DL}
$parentOf \sqsubseteq \neg childOf$
\end{DL}

\begin{gcotable}
property & $\top$ & parentOf & $\neg$childOf & $\top$ & $\sqsubseteq$ \\
\end{gcotable}

syntactic sugar:

\begin{gcotable}
property & $\top$ & parentOf & childOf & $\top$ & $\ne$ \\
\end{gcotable}

\section{Intersection}

\ms{Intersection} (composition, conjunction) corresponds to the requirements \\
\ms{R-15-CONJUNCTION-OF-CLASS-EXPRESSIONS} and \ms{R-16-CONJUNCTION-OF-DATA-} \ms{RANGES}.
DLs allow new concepts and roles to be
built using a variety of different constructors. We distinguish concept and role 
constructors depending on whether concept or role expressions are constructed. In the case of
concepts, one can further separate basic Boolean constructors, role restrictions and nominals/enumerations \cite{Kroetzsch2012}.
Boolean concept constructors provide basic boolean operations that are closely related to
the familiar operations of intersection, union and complement of sets, or to conjunction,
disjunction and negation of logical expressions \cite{Kroetzsch2012}.

\begin{DL}
Mother $\equiv$ Female $\sqcap$ Parent
\end{DL}

Concept inclusions allow us to state that all mothers are female and that
all mothers are parents, but what we really mean is that mothers are exactly the female
parents. DLs support such statements by allowing us to form complex concepts such as
the intersection (also called conjunction)
which denotes the set of individuals that are both female and parents. A complex concept
can be used in axioms in exactly the same way as an atomic concept, e.g., in the
equivalence Mother $\equiv$ Female $\sqcap$ Parent .

\subsection{Simple Example}

\begin{DL}
Female $\sqcap$ Parent
\end{DL}

Complex concept of all individuals which are of the concept \ms{Female} and of the concept \ms{Parent}.

\begin{gcotable}
class & Female $\sqcap$ Parent & - & - & Female, Parent & $\sqcap$ \\
\end{gcotable}

\subsection{Simple Example}

\begin{DL}
Mother $\equiv$ Female $\sqcap$ Parent
\end{DL}

\begin{gcotable}
class & Mother & - & - & Female, Parent & $\sqcap$ \\
\end{gcotable}

\section{Disjunction}

\ms{Disjunction} of classes or data ranges corresponds to the requirements \\
\ms{R-17-DISJUNCTION-OF-CLASS-EXPRESSIONS} and \ms{R-18-DISJUNCTION-OF-} \\
\ms{DATA-RANGES}.
Synonyms are union and inclusive or.

\subsection{Simple Example}

\begin{DL}
Father $\sqcup$ Mother $\sqcup$ Child
\end{DL}

\begin{gcotable}
class & Father $\sqcup$ Mother $\sqcup$ Child & - & - & Father, Mother, Child & $\sqcup$ \\
\end{gcotable}

\subsection{Simple Example}

\begin{DL}
Parent $\equiv$ Father $\sqcup$ Mother
\end{DL}

\begin{gcotable}
class & Parent & - & - & Father, Mother & $\sqcup$ \\
\end{gcotable}

\section{Negation}

\ms{Negation} (complement) corresponds to the requirements \ms{R-19-NEGATION-OF-} \ms{CLASS-EXPRESSIONS} and \ms{R-20-NEGATION-OF-DATA-RANGES}.

\subsection{Simple Example}

\begin{DL}
$ \neg Married $ \\ 
\end{DL}

Set of all individuals that are not married.

\begin{gcotable}
class & $\neg$ Married & - & - & Married & $\neg$ \\
\end{gcotable}

\subsection{Complex Example}

\begin{DL}
Female $\sqcap$ $\neg$ Married \\
\end{DL}

All female individuals that are not married.

\begin{gcotable}
class & $\neg$ Married & - & - & Married & $\neg$ \\
class & Female $\sqcap$ $\neg$ Married & - & - & Female, $\neg$ Married & $\sqcap$ \\
\end{gcotable}

\section{Disjoint Classes}

\ms{Disjoint Classes} corresponds to the requirement \ms{R-7-DISJOINT-CLASSES}.
A disjoint classes axiom states that all of the classes are pairwise disjoint; 
that is, no individual can be at the same time an instance of these classes. 

\subsection{Simple Example}

Individuals cannot be male and female at the same time:

\begin{DL}
Male $\sqcap$ Female $\sqsubseteq$ $\perp$\\
\end{DL}

or alternatively:

\begin{DL}
$Male \sqsubseteq \neg Female$
\end{DL}

\begin{gcotable}
class & Male $\sqcap$ Female & - & - & Male, Female & $\sqcap$  \\
class & $\top$ & - & - & Male $\sqcap$ Female, $\perp$ & $\sqsubseteq$ \\
\end{gcotable}

syntactic sugar:

\begin{gcotable}
class & Male & - & - & Female & $\ne$ \\
class & Female & - & - & Male & $\ne$ \\
\end{gcotable}

\subsection{Simple Example}

One can either be a hologram or a human, but not both:

\begin{DL}
Hologram $\sqcap$ Human $\sqsubseteq$ $\perp$\\
\end{DL}

or alternatively:

\begin{DL}
$Hologram \sqsubseteq \neg Human$
\end{DL}

\begin{gcotable}
class & Hologram $\sqcap$ Human & - & - & Hologram, Human & $\sqcap$  \\
class & $\top$ & - & - & Hologram $\sqcap$ Human, $\perp$ & $\sqsubseteq$ \\
\end{gcotable}

syntactic sugar:

\begin{gcotable}
class & Hologram & - & - & Human & $\ne$ \\
class & Human & - & - & Hologram & $\ne$ \\
\end{gcotable}

\section{Existential Quantifications}

\ms{Existential Quantification on Properties} conforms to the requirement \\
\ms{R-86-EXISTENTIAL-QUANTIFICATION-ON-PROPERTIES}.
In DL terminology the \ms{existential quantification} is also called \ms{existential restriction}.
An existential class expression consists of a property expression and a class expression or a data range, and it contains all those individuals that are connected by the property expression to an individual that is an instance of the class expression or to literals that are in the data range.  

\subsection{Simple Example}

\begin{DL}
$\exists$ parentOf . $\top$
\end{DL}

Complex concept that describes the set
of individuals that are parents of at least one individual (instance of $\top$). 

\begin{gcotable}
property & $\exists$ parentOf . $\top$ & parentOf & - & $\top$ & $\exists$ \\
\end{gcotable}

\subsection{Simple Example}

\begin{DL}
$\exists$ parentOf . Female
\end{DL}

The complex concept describes those individuals that are parents of at least one
female individual, i.e., those that have a daughter.

\begin{gcotable}
property & $\exists$ parentOf . Female & parentOf & - & Female & $\exists$ \\
\end{gcotable}

\subsection{Simple Example}

\begin{DL}
Parent $\equiv$ $\exists$ parentOf . $\top$
\end{DL}

\begin{gcotable}
property & Parent & parentOf & - & $\top$ & $\exists$ \\
\end{gcotable}

\subsection{Simple Example}

\begin{DL}
ParentOfSon $\equiv$ $\exists$ parentOf . Male
\end{DL}

\begin{gcotable}
property & ParentOfSon & parentOf & - & Male & $\exists$ \\
\end{gcotable}

\section{Universal Quantifications}

\ms{Universal Quantification on Properties} corresponds to the requirement \ms{R-91-UNIVERSAL-QUANTIFICATION-ON-PROPERTIES}, 
which is also called \ms{value} \ms{restriction} in DL terminology.

\begin{center}
\begin{DL} 
$(\forall R.C)^\mathcal{I}= \{a \in \Delta^\mathcal{I}\mid (a,b) \in R^\mathcal{I} \rightarrow b \in C^\mathcal{I} \}$ where $\cdot^\mathcal{I}$ is an interpretation function, \\
$\Delta^\mathcal{I}$ is the domain, $a,b \in \Delta^\mathcal{I}$ are individuals $C$ is a concept, $R$ is a role. 
\end{DL}
\end{center}

\subsection{Simple Example}

\begin{DL}
$\forall$ parentOf.Female
\end{DL}

The set of individuals all of whose children are female also includes those that have no children at all.

\begin{gcotable}
property & $\forall$ parentOf . Female & parentOf & - & Female & $\forall$ \\
\end{gcotable}

\section{Property Domains}

\ms{Property Domain} corresponds to the requirements \ms{R-25-OBJECT-PROPERTY-DOMAIN} and \ms{R-26-DATA-PROPERTY-DOMAIN}.
The constraint restricts the domain of object and data properties.
In DL terminology this constraint is also called domain restrictions on roles.
The purpose is to declare that a given property is associated with a class, e.g. to populate input forms with appropriate widgets but also constraint checking. In OO terms this is the declaration of a member, field, attribute or association. 
$\exists R. \top \sqsubseteq C$ is the object property restriction where $R$ is the object property (role) whose domain is restricted to concept $C$.

\subsection{Simple Example}

\begin{DL}
$\exists$ sonOf . $\top$ $\sqsubseteq$ Male 
\end{DL}

Restricts the domain of \ms{sonOf} to male individual.

syntactic sugar:

\begin{gcotable}
property & $\top$ & sonOf & - & Male & domain \\
\end{gcotable}

\section{Property Ranges}

\ms{Property Range} corresponds to the requirements \ms{R-28-OBJECT-PROPERTY-RANGE} and \ms{R-35-DATA-PROPERTY-RANGE}.
This constraint restricts the range of object and data properties
In DL terminology it is also called range restrictions on roles.
$\top \sqsubseteq \forall R . C$ is the range restriction to the object property $R$ (restricted by the concept $C$).  

\subsection{Simple Example}

\begin{DL}
$\top$ $\sqsubseteq$ $\forall$ sonOf . Parent \\
\end{DL}

equivalent to:

\begin{DL}
$\exists sonOf\sp{\overline{\ }}.\top \sqsubseteq Person$
\end{DL}

Restricts the range of \ms{sonOf} to parents.

\begin{ex}
# OWL 2:
sonOf rdfs:range Parent . 
\end{ex}

\begin{ex}
# DSP:
:hasDogRange
        a dsp:DescriptionTemplate ; 
        dsp:resourceClass owl:Thing ; 
        dsp:statementTemplate [
            a dsp:NonLiteralStatementTemplate ;
            dsp:property sonOf ; 
            dsp:nonLiteralConstraint [ 
                a dsp:NonLiteralConstraint ;
                dsp:valueClass Parent ] ] .
\end{ex}

syntactic sugar:

\begin{gcotable}
property & $\top$ & sonOf & - & Parent & range \\
\end{gcotable}

\section{Class-Specific Property Range}

\ms{Class-Specific Property Range} corresponds to the requirements
\ms{R-29-CLASS-} \ms{SPECIFIC-RANGE-OF-RDF-OBJECTS} and \ms{R-36-CLASS-SPECIFIC-RANGE-OF-RDF-} \ms{LITERALS}.
The constraint restricts the range of object and data properties for individuals within a specific context (e.g. class, application profile).
The values of each member property of a class may be limited by their value type, such as xsd:string or Person. 

\subsection{Simple Example}

Only men can have \ms{sonOf} relationships to parents:

\begin{DL}
$\neg$Man $\sqsubseteq$ $\neg\exists$ sonOf.Parent
\end{DL}


\subsection{Simple Example}

Only vulcans can have friendship relationships to cardassians

\begin{DL}
$\neg$Vulcan $\sqsubseteq$ $\neg\exists$ friendOf.Carsassian
\end{DL}

\section{Minimum Unqualified Cardinality Restrictions}

\ms{Minimum Unqualified Cardinality Restrictions on Properties}
corresponds to the requirements
\ms{R-81-MINIMUM-UNQUALIFIED-CARDINALITY-ON-PROPERTIES} and
\ms{R-211-CARDINALITY-CONSTRAINTS}.
$\leq n R. \top$ is the minimum unqualified cardinality restriction where $n \in \mathbb{N}$ (written $\leq  n R$ in short).
A minimum cardinality restrictions contains all those individuals that are connected by a property to at least n different individuals/literals 
that are instances of a particular class or data range. If the class is missing, it is taken to be owl:Thing. 
If the data range is missing, it is taken to be rdfs:Literal.
For unqualified cardinality restrictions, classes respective data ranges are not stated.

\subsection{Simple Example}

\begin{ex}
# ShEx:
:Jedi {
    :attitude ('good') }
:JediStudent {
    & :Jedi ,
    :studentOf {}{1} }
:JediMaster {
    & :Jedi ,
    :mentorOf {}{1,2} }
:SuperJediMaster {
    & :Jedi ,
    :mentorOf {}{3,} }
\end{ex}

\begin{itemize}
	\item Jedis have the attitude 'good'
	\item Jedi students are students of exactly 1 resource
	\item Jedi masters are mentoring at least 1 and at most 2 resources
	\item Super Jedi masters are mentoring at least 3 resources  
\end{itemize}

\begin{ex}
# data:
:Yoda 
    :attitude 'good' ;
    :mentorOf :MaceWindu , :Obi-Wan , :Luke .
:MaceWindu
    :attitude 'good' ;
    :studentOf :Yoda .
:Obi-Wan 
    :attitude 'good' ;
    :studentOf :Yoda ;
    :mentorOf :Anakin .
:Anakin
    :attitude 'good' ; 
    :studentOf :Obi-Wan .
:Luke
    :attitude 'good' ;
    :studentOf :Yoda .
\end{ex}

\begin{ex}
# Individuals matching the ’:SuperJediMaster’ data shape:
:Yoda 
\end{ex}

\begin{DL}
$Jedi \sqsubseteq \exists attitude.\{good\} $\\
$JediStudent \sqsubseteq Jedi \sqcap \geq1 studentOf.\top \sqcap \leq1 studentOf.\top$ \\
$JediMasters \sqsubseteq Jedi \sqcap \geq1 mentorOf.\top \sqcap \leq2 mentorOf.\top $\\
$SuperJediMaster \sqsubseteq Jedi \sqcap  \geq3 mentorOf.\top $
\end{DL}

\subsection{Simple Example}

\begin{DL}
$Captain \sqsubseteq \geq1 commandsVessel.\top $
\end{DL}

Captains command at least one vessel.

\begin{gcotable}
property & Captain & commandsVessel & - & $\top$ & $\geq$ & 1 \\
\end{gcotable}

\section{Minimum Qualified Cardinality Restrictions}

\ms{Minimum Qualified Cardinality Restrictions on Properties} corresponds to the requirements
\ms{R-75-MINIMUM-QUALIFIED-CARDINALITY-ON-PROPERTIES} and \ms{R-211-CARDINALITY-CONSTRAINTS}
A minimum cardinality restrictions contains all those individuals that are connected by a property to at least n different individuals/literals 
that are instances of a particular class or data range. If the class is missing, it is taken to be owl:Thing. 
If the data range is missing, it is taken to be rdfs:Literal.
For qualified cardinality restrictions, classes respective data ranges are stated.
$\geq n R. C$ is a minimum qualified cardinality restriction where $n \in \mathbb{N}$.

\subsection{Simple Example}

\begin{DL}
$\geq$ 2 childOf . Parent
\end{DL}

Set of individuals that are children of at least two parents.

\begin{ex}
# OWL 2:
owl:Thing
    a owl:Restriction ;
    owl:minQualifiedCardinality "2"^^xsd:nonNegativeInteger ;
    owl:onProperty childOf ;
    owl:onClass Parent .
\end{ex}

\begin{gcotable}
property & $\geq$ 2 childOf . Parent & childOf & -  & Parent & $\geq$ & 2 \\
\end{gcotable}

\subsection{Simple Example}

\begin{DL}
foaf:Person $\equiv$ $\geq$ 2 hasName.xsd:string
\end{DL}

\begin{gcotable}
property & foaf:Person & hasName & - & xsd:string & $\geq$ & 2 \\
\end{gcotable}

\subsection{Simple Example}


\begin{ex}
# ShEx:
:Jedi {
    :attitude ('good') }
:JediStudent {
    & :Jedi ,
    :studentOf @:Jedi{1} }
:JediMaster {
    & :Jedi ,
    :mentorOf @:Jedi{1,2} }
:SuperJediMaster {
    & :Jedi ,
    :mentorOf @:Jedi{3,} }
\end{ex}

\begin{itemize}
	\item Jedis have the attitude 'good'
	\item Jedi students are students of exactly 1 Jedi
	\item Jedi masters are mentoring at least 1 and at most 2 Jedis
	\item Super Jedi masters are mentoring at least 3 Jedis
\end{itemize}

\begin{ex}
# data:
:Yoda 
    :attitude 'good' ;
    :mentorOf :MaceWindu , :Obi-Wan , :Luke .
:MaceWindu
    :attitude 'good' ;
    :studentOf :Yoda .
:Obi-Wan 
    :attitude 'good' ;
    :studentOf :Yoda ;
    :mentorOf :Anakin .
:Anakin
    :attitude 'good' ; 
    :studentOf :Obi-Wan .
:Luke
    :attitude 'good' ;
    :studentOf :Yoda .
\end{ex}

\begin{ex}
# Individuals matching the ’:SuperJediMaster’ data shape:
:Yoda 

# Individuals matching the ’:JediMaster’ data shape:
:Obi-Wan
\end{ex}

\noindent $mentorOf$ and $studentOf$ are taken to be inverse properties:\\

\begin{DL}
$Jedi \sqsubseteq \exists attitude.\{good\} $\\
$JediStudent \sqsubseteq Jedi \sqcap \geq1 studentOf.Jedi \sqcap \leq1 studentOf.Jedi$ \\
$JediMasters \sqsubseteq Jedi \sqcap \geq1 mentorOf.Jedi \sqcap \leq2 mentorOf.Jedi $\\
$SuperJediMaster \sqsubseteq Jedi \sqcap  \geq3 mentorOf.Jedi $
\end{DL}

\subsection{Complex Example}

\begin{DL}
Father2Daughters $\equiv$ Man $\sqcap$ ($\geq$ 2 hasChild.Woman)
\end{DL}

\begin{gcotable}
property & $\geq$ 2 hasChild.Woman & hasChild & - & Woman & $\geq$ & 2 \\
class & Father2Daughters & - & - & Man, $\geq$ 2 hasChild.Woman & $\sqcap$ & - \\
\end{gcotable}

\subsection{Simple Example}

\begin{DL}
$Captain \sqsubseteq \geq1 commandsVessel . Vessel $
\end{DL}

\begin{gcotable}
property & Captain & commandsVessel & - & Vessel & $\geq$ & 1 \\
\end{gcotable}

\subsection{Complex Example}

\begin{DL}
$FederationCaptain \sqsubseteq Federation \sqcap \geq1 commandsVessel . Vessel $
\end{DL}

\begin{gcotable}
property & $\geq$1 commandsVessel . Vessel & commandsVessel & - & Vessel & $\geq$ & 1 \\
class & FederationCaptain & - & - & Federation, $\geq$1 commandsVessel . Vessel & $\sqcap$ & - \\
\end{gcotable}

\section{Maximum Unqualified Cardinality Restrictions}

\ms{Maximum Unqualified Cardinality Restrictions on Properties} corresponds to the requirements 
\ms{R-82-MAXIMUM-UNQUALIFIED-CARDINALITY-ON-PROPERTIES} and \ms{R-211-CARDINALITY-CONSTRAINTS}.
A maximum cardinality restriction contains all those individuals that are connected by a property to at most n different individuals/literals that are instances of a particular class or data range. If the class is missing, it is taken to be owl:Thing. If the data range is not present, it is taken to be rdfs:Literal.
Unqualified means that the class respective the data range is not stated. 
$\geq n R. \top$ is a maximum unqualified cardinality restriction where $n \in \mathbb{N}$ (written $\geq  n R$ in short).
\subsection{Simple Example}

\begin{ex}
# ShEx:
:Jedi {
    :attitude ('good') }
:JediStudent {
    & :Jedi ,
    :studentOf {}{1} }
:JediMaster {
    & :Jedi ,
    :mentorOf {}{1,2} }
:SuperJediMaster {
    & :Jedi ,
    :mentorOf {}{3,} }
\end{ex}

\begin{itemize}
	\item Jedis have the attitude 'good'
	\item Jedi students are students of exactly 1 resource
	\item Jedi masters are mentoring at least 1 and at most 2 resources
	\item Super Jedi masters are mentoring at least 3 resources  
\end{itemize}
\begin{ex}
# data:
:Yoda 
    :attitude 'good' ;
    :mentorOf :MaceWindu , :Obi-Wan , :Luke .
:MaceWindu
    :attitude 'good' ;
    :studentOf :Yoda .
:Obi-Wan 
    :attitude 'good' ;
    :studentOf :Yoda ;
    :mentorOf :Anakin .
:Anakin
    :attitude 'good' ; 
    :studentOf :Obi-Wan .
:Luke
    :attitude 'good' ;
    :studentOf :Yoda .
\end{ex}

\begin{ex}
# Individuals matching the ’:JediMaster’ data shape:
:Obi-Wan
\end{ex}

\begin{DL}
$Jedi \sqsubseteq \exists attitude.\{good\} $\\
$JediStudent \sqsubseteq Jedi \sqcap \geq1 studentOf.\top \sqcap \leq1 studentOf.\top$ \\
$JediMasters \sqsubseteq Jedi \sqcap \geq1 mentorOf.\top \sqcap \leq2 mentorOf.\top $\\
$SuperJediMaster \sqsubseteq Jedi \sqcap  \geq3 mentorOf.\top $
\end{DL}

\section{Maximum Qualified Cardinality Restrictions}

\ms{Maximum Qualified Cardinality Restrictions on Properties} corresponds to the requirements
\ms{R-76-MAXIMUM-QUALIFIED-CARDINALITY-ON-PROPERTIES} and
\ms{R-211-CARDINALITY-CONSTRAINTS}.
A maximum cardinality restriction contains all those individuals that are connected by a property to at most n different individuals/literals that are instances of a particular class or data range. If the class is missing, it is taken to be owl:Thing. If the data range is not present, it is taken to be rdfs:Literal.
Qualified means that the class respective the data range is stated. 
$\leq n R. C$ is a maximum qualified cardinality restriction where $n \in \mathbb{N}$.

\subsection{Simple Example}

\begin{DL}
$\leq$  2 childOf . Parent
\end{DL}

Set of individuals that are children of at most two parents.

\begin{gcotable}
property & $\leq$  2 childOf . Parent & childOf & - & Parent & $\leq$ & 2 \\
\end{gcotable}

\subsection{Simple Example}

\begin{ex}
# ShEx:
:Jedi {
    :attitude ('good') }
:JediStudent {
    & :Jedi ,
    :studentOf @:Jedi{1} }
:JediMaster {
    & :Jedi ,
    :mentorOf @:Jedi{1,2} }
:SuperJediMaster {
    & :Jedi ,
    :mentorOf @:Jedi{3,} }
\end{ex}

\begin{itemize}
	\item Jedis have the attitude 'good'
	\item Jedi students are students of exactly 1 Jedi
	\item Jedi masters are mentoring at least 1 and at most 2 Jedis
	\item Super Jedi masters are mentoring at least 3 Jedis
\end{itemize}

\begin{ex}
# data:
:Yoda 
    :attitude 'good' ;
    :mentorOf :MaceWindu , :Obi-Wan , :Luke .
:MaceWindu
    :attitude 'good' ;
    :studentOf :Yoda .
:Obi-Wan 
    :attitude 'good' ;
    :studentOf :Yoda ;
    :mentorOf :Anakin .
:Anakin
    :attitude 'good' ; 
    :studentOf :Obi-Wan .
:Luke
    :attitude 'good' ;
    :studentOf :Yoda .
\end{ex}

\begin{ex}
# Individuals matching the ’:JediMaster’ data shape:
:Obi-Wan
\end{ex}


\begin{DL}
$Jedi \sqsubseteq \exists attitude.\{good\} $\\
$JediStudent \sqsubseteq Jedi \sqcap \geq1 studentOf.Jedi \sqcap \leq1 studentOf.Jedi$ \\
$JediMasters \sqsubseteq Jedi \sqcap \geq1 mentorOf.Jedi \sqcap \leq2 mentorOf.Jedi $\\
$SuperJediMaster \sqsubseteq Jedi \sqcap  \geq3 mentorOf.Jedi $
\end{DL}

\section{Exact Unqualified Cardinality Restrictions}

\ms{Exact Unqualified Cardinality Restrictions on Properties} corresponds to the requirements
\ms{R-80-EXACT-UNQUALIFIED-CARDINALITY-ON-PROPERTIES} and
\ms{R-211-CARDINALITY-CONSTRAINTS}.
An exact cardinality restriction contains all those individuals that are connected by a property to exactly n different individuals that are instances of a particular class or data range. 
If the class is missing, it is taken to be owl:Thing. 
If the data range is not present, it is taken to be rdfs:Literal.
Unqualified means that the class respective data range is not stated. 
$\geq n R. \top \sqcap \leq n R. \top $ is an exact unqualified cardinality restriction where $n \in \mathbb{N}$.

\subsection{Simple Example}


\begin{ex}
# ShEx:
:JediStudent {
    :studentOf {}{1} }
\end{ex}

\begin{ex}
# ReSh:
:JediStudent a rs:ResourceShape ;
    rs:property [
        rs:name "studentOf" ;
        rs:propertyDefinition :studentOf ;
        rs:valueShape [ a rs:ResourceShape] ;
        rs:occurs rs:Exactly-one ; ] .
\end{ex}

\begin{itemize}
	\item Jedis have the attitude 'good'
	\item Jedi students are students of exactly 1 resource
\end{itemize}

\begin{ex}
# data:
:Yoda 
    :attitude 'good' ;
    :mentorOf :MaceWindu , :Obi-Wan , :Luke .
:MaceWindu
    :attitude 'good' ;
    :studentOf :Yoda .
:Obi-Wan 
    :attitude 'good' ;
    :studentOf :Yoda ;
    :mentorOf :Anakin .
:Anakin
    :attitude 'good' ; 
    :studentOf :Obi-Wan .
:Luke
    :attitude 'good' ;
    :studentOf :Yoda .
\end{ex}

\begin{ex}
# Individuals matching the ’:JediStudent’ data shape:
:MaceWindu :Obi-Wan :Anakin :Luke
\end{ex}

\begin{DL}
$JediStudent \sqsubseteq Jedi \sqcap \geq1 studentOf.\top \sqcap \leq1 studentOf.\top$ \\
\end{DL}

\section{Exact Qualified Cardinality Restrictions}

\ms{Exact Qualified Cardinality Restrictions on Properties} corresponds to the requirements
\ms{R-74-EXACT-QUALIFIED-CARDINALITY-ON-PROPERTIES} and \\
\ms{R-211-CARDINALITY-CONSTRAINTS}.
An exact cardinality restriction contains all those individuals that are connected by a property to exactly n different individuals that are instances of a particular class or data range. 
If the class is missing, it is taken to be owl:Thing. 
If the data range is not present, it is taken to be rdfs:Literal.
Qualified means that the class respective data range is stated. 
$\geq n R. C \sqcap \leq n R. C $ is an exact qualified cardinality restriction where $n \in \mathbb{N}$.

\subsection{Simple Example}


\begin{ex}
# ShEx:
:Jedi {
    :attitude ('good') }
:JediStudent {
    :studentOf @:Jedi{1} }
\end{ex}

\begin{ex}
# ReSh:
:Jedi a rs:ResourceShape ;
    rs:property [
        rs:name "attitude" ;
        rs:propertyDefinition :attitude ;
        rs:allowedValue "good" ;
        rs:occurs rs:Exactly-one ;
    ] .
:JediStudent a rs:ResourceShape ;
    rs:property [
        rs:name "studentOf" ;
        rs:propertyDefinition :studentOf ;
        rs:valueShape :Jedi ;
        rs:occurs rs:Exactly-one ;
    ] .
\end{ex}

\begin{itemize}
	\item Jedis have the attitude 'good'
	\item Jedi students are students of exactly 1 Jedi
\end{itemize}

\begin{ex}
# data:
:Yoda 
    :attitude 'good' ;
    :mentorOf :MaceWindu , :Obi-Wan , :Luke .
:MaceWindu
    :attitude 'good' ;
    :studentOf :Yoda .
:Obi-Wan 
    :attitude 'good' ;
    :studentOf :Yoda ;
    :mentorOf :Anakin .
:Anakin
    :attitude 'good' ; 
    :studentOf :Obi-Wan .
:Luke
    :attitude 'good' ;
    :studentOf :Yoda .
\end{ex}

\begin{ex}
# Individuals matching the ’:JediStudent’ data shape:
:MaceWindu :Obi-Wan :Anakin :Luke
\end{ex}

\begin{DL}
$Jedi \sqsubseteq \exists attitude.\{good\} $\\
$JediStudent \sqsubseteq Jedi \sqcap \geq1 studentOf.Jedi \sqcap \leq1 studentOf.Jedi$ \\
\end{DL}

\subsection{Complex Example}

\begin{DL}
Person $\sqsubseteq$ $\geq$ 2 childOf . Parent $\sqcap$ $\leq$  2 childOf . Parent \\
\end{DL}

Every person is a child of exactly two parents.

\begin{gcotable}
property & $\geq$ 2 childOf . Parent & childOf & - & Parent & $\geq$ & 2 \\
property & $\leq$ 2 childOf . Parent & childOf & - & Parent & $\leq$ & 2 \\
class & Person & - & - & $\geq$ 2 childOf . Parent, $\leq$ 2 childOf . Parent & $\sqcap$ & - \\
\end{gcotable}

syntactic sugar:

\begin{gcotable}
property & Person & childOf & - & Parent & = & 2 \\
\end{gcotable}

\section{Inverse Object Properties}

\ms{Inverse Object Properties} corresponds to the requirement
\ms{R-56-INVERSE-} \ms{OBJECT-PROPERTIES}.
In many cases properties are used bi-directionally and then accessed in the inverse direction, e.g. parent $\equiv$ child$^{-}$. There should be a way to declare value type, cardinality etc of those inverse relations without having to declare a new property URI. 
The object property OP1 is an inverse of the object property OP2. 
Thus, if an individual x is connected by OP1 to an individual y, then y is also connected by OP2 to x, and vice versa.

\subsection{Simple Example}

\begin{DL}
$parentOf \equiv childOf^{-}$ \\
\end{DL}

\begin{gcotable}
property & $\top$ & parentOf & childOf & - & inverse \\
\end{gcotable}

%

\subsection{Simple Example}

\begin{DL}
captainOf $\equiv$ hasCaptain$^{-}$ \\
\end{DL}

\begin{gcotable}
property & $\top$ & captainOf & hasCaptain & - & inverse \\
\end{gcotable}

%
%
%

\section{Transitive Object Properties}

\ms{Transitive Object Properties} corresponds to the requirement \\
\ms{R-63-TRANSITIVE-OBJECT-PROPERTIES}.
Transitivity is a special form of complex role inclusion.
An object property transitivity axiom states that the object property is transitive — that is, if an individual x is connected by the object property to an individual y that is connected by the object property to an individual z, then x is also connected by the object property to z.

\subsection{Simple Example}

\begin{DL}
ancestorOf $\circ$ ancestorOf $\sqsubseteq$ ancestorOf
\end{DL}

\begin{gcotable}
property & $\top$ & ancestorOf, ancestorOf & ancestorOf & - & $\sqsubseteq$ \\
\end{gcotable}

syntactic sugar:

\begin{gcotable}
property & $\top$ & ancestorOf & - & - & transitive \\
\end{gcotable}

\section{Symmetric Object Properties}

\ms{Symmetric Object Properties} corresponds to the requirement \\
\ms{R-61-SYMMETRIC-OBJECT-PROPERTIES}.
A role is symmetric if it is equivalent to its own inverse \cite{Kroetzsch2012}.
An object property symmetry axiom states that the object property expression OPE is symmetric - that is, if an individual x is connected by OPE to an individual y, then y is also connected by OPE to x. 	

\subsection{Simple Example}

The property \ms{marriedTo} is symmetric:

\begin{DL}
$marriedTo \equiv marriedTo^{-}$ \\
\end{DL}

\begin{gcotable}
property & $\top$ & marriedTo & marriedTo & - & inverse \\
\end{gcotable}

syntactic sugar:

\begin{gcotable}
property & $\top$ & marriedTo & - & - & symmetric \\
\end{gcotable}

\section{Asymmetric Object Properties}

\ms{Asymmetric Object Properties} corresponds to the requirement \\
\ms{R-62-ASYMMETRIC-OBJECT-PROPERTIES}.
A role is asymmetric if it is disjoint from its own inverse \cite{Kroetzsch2012}.
An object property asymmetry axiom AsymmetricObjectProperty( OPE ) states that the object property expression OPE is asymmetric - that is, if an individual x is connected by OPE to an individual y, then y cannot be connected by OPE to x. 

\subsection{Simple Example}

The property \ms{parentOf} is asymmetric:

\begin{DL}
$parentOf \sqsubseteq \neg parentOf^{-}$ 
\end{DL}

alternatively:

\begin{DL}
$parentOf \sqcap parentOf^{-} \sqsubseteq \bot$ 
\end{DL}

alternatively:

\begin{DL}
disjoint(parentOf, parentOf$^{-}$)
\end{DL}

\begin{gcotable}
property & $\top$ & $parentOf^{-}$ & parentOf & - & inverse \\
property & $\top$ & parentOf & $parentOf^{-}$ & - & $\ne$ \\
\end{gcotable}

syntactic sugar:

\begin{gcotable}
property & $\top$ & parentOf & - & - & asymmetric \\
\end{gcotable}

\subsection{Simple Example}

Child parent relations (dbo:child) cannot be symmetric.

\subsection{Simple Example}

Person birth place relations (dbo:birthPlace) cannot be symmetric.

\section{Class-Specific Reflexive Object Properties}

Using DL terminology \ms{Class-Specific Reflexive Object Properties} is called local reflexivity - a set of individuals (of a specific class) that are related to themselves via a given role \cite{Kroetzsch2012}.

\subsection{Simple Example}

\begin{DL}
TalkingToThemselves $\sqsubseteq$ $\exists$ talksTo .Self. 
\end{DL}

Set of individuals that are talking to themselves.

\begin{gcotable}
property & TalkingToThemselves & talksTo & - & Self & $\exists$ \\
\end{gcotable}

syntactic sugar:

\begin{gcotable}
property & $\exists$ talksTo .Self & talksTo & - & - & reflexive \\
\end{gcotable}

\section{Reflexive Object Properties}

\ms{Reflexive Object Properties} (reflexive roles, global reflexivity in DL) corresponds to the requirement \ms{R-59-REFLEXIVE-OBJECT-PROPERTIES}.
Global reflexivity can be expressed by imposing local reflexivity on the top concept \cite{Kroetzsch2012}.

\subsection{Simple Example}

Each individual knows itself:

\begin{DL}
$\top$ $\sqsubseteq$ $\exists$ knows . Self
\end{DL}

\begin{gcotable}
property & $\exists$ knows . Self & knows & - & Self & $\exists$ \\
class & $\top$ & - & - & $\top$, $\exists$ knows . Self & $\sqsubseteq$ \\
\end{gcotable}

syntactic sugar:

\begin{gcotable}
property & $\top$ & knows & - & - & reflexive \\
\end{gcotable}

\section{Irreflexive Object Properties}

\ms{Irreflexive Object Properties} (irreflexive roles in DL) corresponds to the requirement \ms{R-60-IRREFLEXIVE-OBJECT-PROPERTIES}.
A role is irreflexive if it is never locally reflexive \cite{Kroetzsch2012}.
An object property irreflexivity axiom IrreflexiveObjectProperty( OPE ) states that the object property expression OPE is irreflexive - that is, no individual is connected by OPE to itself. 

\subsection{Simple Example}

\begin{DL}
$\top$ $\sqsubseteq$ $\neg$ $\exists  marriedTo . Self$ 
\end{DL}

alternatively:

\begin{DL}
$\exists marriedTo . Self \sqsubseteq \bot$  
\end{DL}

alternatively:

\begin{DL}
$marriedTo \sqcap marriedTo^{-} \sqsubseteq \bot$ (without special self-concept Self)
\end{DL}

\begin{gcotable}
property & $\exists$ marriedTo . Self & knows & - & Self & $\exists$ \\
class & $\neg$ $\exists$ marriedTo . Self & - & - & $\exists$ marriedTo . Self & $\neg$ \\
class & $\top$ & - & - & $\top$, $\neg$ $\exists$ marriedTo . Self & $\sqsubseteq$ \\
\end{gcotable}

syntactic sugar:

\begin{gcotable}
property & $\top$ & marriedTo & - & - & irreflexive \\
\end{gcotable}

\subsection{Simple Example}

A resource cannot be its own parent (dbo:parent).

\subsection{Simple Example}

A resource cannot be its own child (dbo:child).

\subsection{Class-Specific Irreflexive Object Properties}

A property is \emph{irreflexive} if it is never locally reflexive \cite{Kroetzsch2012}.
An object property irreflexivity axiom states that the object property \emph{OP} is irreflexive - that is, no individual is connected by \emph{OP} to itself.
\emph{Class-Specific Irreflexive Object Properties} are object properties which are irreflexive within a given context, e.g. a class. 

\subsection{Simple Example}

\emph{Persons} cannot be married to themselves.

\begin{DL}
Person $\sqsubseteq$ $\neg$ $\exists  marriedTo . Self$ 
\end{DL}

\begin{gcotable}
property & $\exists$ marriedTo . Self & knows & - & Self & $\exists$ \\
class & $\neg$ $\exists$ marriedTo . Self & - & - & $\exists$ knows . Self & $\neg$ \\
class & Person & - & - & Person, $\neg$ $\exists$ marriedTo . Self & $\sqsubseteq$ \\
\end{gcotable}

syntactic sugar:

\begin{gcotable}
property & Person & marriedTo & - & - & irreflexive \\
\end{gcotable}

\section{Context-Specific Property Groups}

\ms{Context-Specific Property Groups} corresponds to the requirement \\
\ms{R-66-PROPERTY-GROUPS} to group data and object properties.

\subsection{Simple Example}

\begin{ex}
# ShEx:
<Human> { 
    (  
        foaf:name xsd:string ,
        foaf:givenName xsd:string 
    ) ,
    (
        foaf:mbox IRI ,
        foaf:homepage foaf:Document
    ) }
\end{ex}

\begin{itemize}
	\item 1. group: 1 foaf:name (range: xsd:string) and 1 foaf:givenName (range: xsd:string) and (,)
	\item 2. group: 1 foaf:mbox (range: IRI) and 1 foaf:homepage (range: foaf:Document) 
\end{itemize}


\begin{DL}
Human $\equiv$ M $\sqcap$ N \\
M $\equiv$ C $\sqcap$ F \\
C $\sqsubseteq$ $\geq$ 1 name . string $\sqcap$ $\leq$ 1 name . string \\
F $\sqsubseteq$ $\geq$ 1 givenName . string $\sqcap$ $\leq$ 1 givenName . string \\
N $\equiv$ I $\sqcap$ L \\
I $\sqsubseteq$ $\geq$ 1 mbox . string $\sqcap$ $\leq$ 1 mbox . string \\
L $\sqsubseteq$ $\geq$ 1 homepage . Document $\sqcap$ $\leq$ 1 homepage . Document \\
\end{DL}

\begin{gcotable}
class & Human & - & - & M, N & $\sqcap$ & - \\
\hline
class & M & - & - & C, F & $\sqcap$ & - \\
class & C & - & - & A, B & $\sqcap$ & - \\
property & A & foaf:name & - & string & $\geq$ & 1 \\
property & B & foaf:name & - & string & $\leq$ & 1 \\
class & F & - & - & D, E & $\sqcap$ & - \\
property & D & foaf:givenName & - & string & $\geq$ & 1 \\
property & E & foaf:givenName & - & string & $\leq$ & 1 \\
\hline
class & N & - & - & I, L & $\sqcap$ & - \\
class & I & - & - & G, H & $\sqcap$ & - \\
property & G & foaf:mbox & - & string & $\geq$ & 1 \\
property & H & foaf:mbox & - & string & $\leq$ & 1 \\
class & L & - & - & J, K & $\sqcap$ & - \\
property & J & foaf:homepage & - & foaf:Document & $\geq$ & 1 \\
property & K & foaf:homepage & - & foaf:Document & $\leq$ & 1 \\
\end{gcotable}

\section{Context-Specific Exclusive OR of Properties}

\ms{Context-Specific Exclusive OR of Properties} correponds to the requirement
\ms{R-11-CONTEXT-SPECIFIC-EXCLUSIVE-OR-OF-PROPERTIES}.
Exclusive or is a logical operation that outputs true whenever both inputs differ (one is true, the other is false).
This constraint is generally expressed in DL as follows:

\begin{DL}
$C \sqsubseteq (\neg A \sqcap B) \sqcup (A \sqcap \neg B)$ \\
\end{DL}

and alternatively:

\begin{DL}
$C \sqsubseteq (A \sqcup B) \sqcap \neg(A \sqcap B)$ \\
\end{DL}

\subsection{Simple Example}

\begin{ex}
# ShEx:
<Human> { (  
    foaf:name xsd:string | 
    foaf:givenName xsd:string ) }
\end{ex}

\begin{ex}
# OWL 2:
Human owl:disjointUnionOf ( :CC1 :CC2 ) . 

CC1 rdfs:subClassOf [
    a owl:Restriction ;
    owl:onProperty foaf:name ;
    owl:someValuesFrom xsd:string ] .
CC2 rdfs:subClassOf [
    a owl:Restriction ;
    owl:onProperty foaf:givenName ;
    owl:someValuesFrom xsd:string ] .
\end{ex}

\begin{itemize}
	\item 1 foaf:name (range xsd:string) XOR 1 foaf:givenName (range xsd:string)
\end{itemize}

\begin{DL}
Human $\sqsubseteq$ ($\neg$ A $\sqcap$ B) $\sqcup$ (A $\sqcap$ $\neg$ B) \\
A $\sqsubseteq$ $\geq$ 1 name . string $\sqcap$ $\leq$ 1 name . string \\
B $\sqsubseteq$ $\geq$ 1 givenName . string $\sqcap$ $\leq$ 1 givenName . string \\
\end{DL}

\begin{gcotable}
class & Human & - & - & $\neg$ A $\sqcap$ B, A $\sqcap$ $\neg$ B & $\sqcup$ \\
\hline
class & A $\sqcap$ $\neg$ B & - & - & A, $\neg$ B & $\sqcap$ \\
class & $\neg$ A $\sqcap$ B & - & - & $\neg$ A, B & $\sqcap$ \\
\hline
class & $\neg$ A & - & - & A & $\neg$ \\
class & A & - & - & $\geq$ 1 name.string, $\leq$ 1 name.string & $\sqcap$ & - \\
property & $\geq$ 1 name.string & name & - & string & $\geq$ & 1 \\
property & $\leq$ 1 name.string & name & - & string & $\leq$ & 1 \\
\hline
class & $\neg$ B & - & - & B & $\neg$ \\
class & B & - & - & $\geq$ 1 givenName.string, $\leq$ 1 givenName.string & $\sqcap$ & - \\
property & $\geq$ 1 givenName.string & givenName & - & string & $\geq$ & 1 \\
property & $\leq$ 1 givenName.string & givenName & - & string & $\leq$ & 1 \\
\end{gcotable}

syntactic sugar:
\begin{gcotable}
property & = 1 name.string & name & - & string & = & 1 \\
property & = 1 givenName.string & givenName & - & string & = & 1 \\
class & Human & - & - & = 1 name.string, = 1 givenName.string & XOR \\
\end{gcotable}

\subsection{Simple Example}

\begin{ex}
# ShEx:
<FeelingForce> { (  
    attackingBySword xsd:boolean | 
    attackingByForce xsd:boolean ) }
\end{ex}

\begin{ex}
# OWL 2 DL:
FeelingForce owl:disjointUnionOf ( CC1 CC2 ) . 
CC1 rdfs:subClassOf [
    a owl:Restriction ;
    owl:onProperty attackingBySword ;
    owl:someValuesFrom xsd:boolean ] .
CC2 rdfs:subClassOf [
    a owl:Restriction ;
    owl:onProperty attackingByForce ;
    owl:someValuesFrom xsd:boolean ] .
\end{ex}

This means that \ms{FeelingForce} individuals are individuals having one \\
\ms{attackingBySword} relationship (range xsd:boolean) or one \ms{attackingByForce} relationship (range xsd:boolean) but not both.

\begin{DL}
A $\equiv$ $\exists$ attackingBySword.xsd:boolean \\
B $\equiv$ $\exists$ attackingByForce.xsd:boolean \\ 
FeelingForce $\sqsubseteq$ ($\neg$ A $\sqcap$ B) $\sqcup$ (A $\sqcap$ $\neg$ B) \\
\end{DL}

\subsection{Simple Example}

\begin{DL}
$Person \sqsubseteq ((Male \sqcap \neg Female) \sqcup (\neg Male \sqcap  Female)) $ 
\end{DL}

\subsection{Complex Example}

\begin{ex}
# ShEx:
<Human> { (  
    foaf:name xsd:string | foaf:givenName xsd:string+ , 
    foaf:familyName xsd:string ) }
\end{ex}

\begin{itemize}
	\item 1 foaf:name (range xsd:string) XOR 1-n foaf:givenName (range xsd:string)
	\item and
	\item 1 foaf:familyName (range xsd:string)
\end{itemize}

Individuals matching the 'Human' data shape:

\begin{ex}
:Han
    foaf:name "Han Solo" ;
    foaf:familyName "Solo" .
:Anakin
    foaf:givenName "Anakin" ;
    foaf:givenName "Darth" ;
    foaf:familyName "Skywalker" .
\end{ex}

Individual not matching the 'Human' data shape:

\begin{ex}
:Anakin
    foaf:name "Anakin Skywalker" ;
    foaf:givenName "Anakin" ;
    foaf:familyName "Skywalker" .
\end{ex}

\section{Context-Specific Inclusive OR of Properties}

\ms{Context-Specific Inclusive OR of Properties} corresponds to the requirement
\ms{R-202-CONTEXT-SPECIFIC-INCLUSIVE-OR-OF-PROPERTIES}.
Inclusive or is a logical connective joining two or more predicates that yields the logical value "true" when at least one of the predicates is true.
The context can be a class, i.e., the constraint applies for individuals of this specific class.

\subsection{Simple Example}

\begin{itemize}
	\item 1 foaf:name (range: xsd:string) OR 1 foaf:givenName (range: xsd:string)
	\item it is also possible that both 1 foaf:name and 1 foaf:givenName are stated
	\item context: class Human
\end{itemize}

\begin{DL}
Human $\sqsubseteq$ A $\sqcup$ B \\
A $\sqsubseteq$ $\geq$ 1 name . string $\sqcap$ $\leq$ 1 name . string \\
B $\sqsubseteq$ $\geq$ 1 givenName . string $\sqcap$ $\leq$ 1 givenName . string \\
\end{DL}

syntactic sugar:
\begin{gcotable}
property & = 1 name.string & name & - & string & = & 1 \\
property & = 1 givenName.string & givenName & - & string & = & 1 \\
class & Human & - & - & = 1 name.string, = 1 givenName.string & $\sqcup$ \\
\end{gcotable}

\section{Context-Specific Exclusive OR of Property Groups}

\ms{Context-Specific Exclusive OR of Property Groups} corresponds to the requirement:
\ms{R-13-DISJOINT-GROUP-OF-PROPERTIES-CLASS-SPECIFIC}.
Exclusive or is a logical operation that outputs true whenever both inputs differ (one is true, the other is false).
Only one of multiple property groups leads to valid data.

\subsection{Simple Example}

\begin{ex}
# ShEx:
<Human> { 
    (  
        foaf:name xsd:string ,
        foaf:givenName xsd:string ) 
    |
    (
        foaf:mbox IRI ,
        foaf:homepage foaf:Document ) }
\end{ex}

\begin{itemize}
  \item 1. group XOR 2. group
	\item 1. group: 1 foaf:name (range: xsd:string) and 1 foaf:givenName (range: xsd:string)
	\item 2. group: 1 foaf:mbox (range: IRI) and 1 foaf:homepage (range: foaf:Document) 
	\item context: class Human
\end{itemize}

\begin{ex}
# valid data:
Thomas
    a Human ;
    foaf:mbox <thomas.bosch@gesis.org> ;
    foaf:homepage <http://purl.org/net/thomasbosch> .
\end{ex}

\begin{ex}
# invalid data:
Thomas
    a Human ;
    foaf:name 'Thomas Bosch' ;
    foaf:givenName 'Thomas' ;
    foaf:mbox <thomas.bosch@gesis.org> ;
    foaf:homepage <http://purl.org/net/thomasbosch> .
\end{ex}

\begin{DL}
Human $\sqsubseteq$ ($\neg$ E $\sqcap$ F) $\sqcup$ (E $\sqcap$ $\neg$ F) \\ 
E $\equiv$ A $\sqcap$ B \\
F $\equiv$ C $\sqcap$ D \\
A $\sqsubseteq$ $\geq$ 1 name . string $\sqcap$ $\leq$ 1 name . string \\
B $\sqsubseteq$ $\geq$ 1 givenName . string $\sqcap$ $\leq$ 1 givenName . string \\
C $\sqsubseteq$ $\geq$ 1 mbox . IRI $\sqcap$ $\leq$ 1 mbox . IRI \\
D $\sqsubseteq$ $\geq$ 1 homepage . Document $\sqcap$ $\leq$ 1 homepage . Document \\
\end{DL}

syntactic sugar:
\begin{gcotable}
class & Human & - & - & E, F & XOR \\
class & E & - & - & = 1 name.string, = 1 givenName.string & $\sqcap$ \\
class & F & - & - & = 1 mbox.IRI, = 1 homepage.Document & $\sqcap$ \\
property & = 1 name.string & name & - & string & = & 1 \\
property & = 1 givenName.string & givenName & - & string & = & 1 \\
property & = 1 mbox.IRI & mbox & - & IRI & = & 1 \\
property & = 1 homepage.Document & homepage & - & Document & = & 1 \\
\end{gcotable}

%
%
%
%

\section{Context-Specific Inclusive OR of Property Groups}

At least one property group must match for individuals of a specific context. 
Context may be a class, a shape, or an application profile.

\subsection{Simple Example}

\begin{itemize}
  \item 1. group OR 2. group
	\item 1. group: 1 foaf:firstName (range: xsd:string) and 1 foaf:lastName (range: xsd:string)
	\item 2. group: 1 foaf:givenName (range: xsd:string) and 1 foaf:familyName (range: xsd:string)
	\item context: class Person
\end{itemize}

\begin{ex}
# valid data:
:Anakin
    a :Person ;
    foaf:firstName 'Anakin' ;
    foaf:lastName 'Skywalker' ;
    foaf:givenName 'Anakin' ;
    foaf:familyName 'Skywalker' .
\end{ex}

\begin{ex}
# invalid data:
:Anakin
    a :Person .
\end{ex}

\begin{DL}
Human $\sqsubseteq$ E $\sqcup$ F \\ 
E $\equiv$ A $\sqcap$ B \\
F $\equiv$ C $\sqcap$ D \\
A $\sqsubseteq$ $\geq$ 1 name . string $\sqcap$ $\leq$ 1 name . string \\
B $\sqsubseteq$ $\geq$ 1 givenName . string $\sqcap$ $\leq$ 1 givenName . string \\
C $\sqsubseteq$ $\geq$ 1 mbox . IRI $\sqcap$ $\leq$ 1 mbox . IRI \\
D $\sqsubseteq$ $\geq$ 1 homepage . Document $\sqcap$ $\leq$ 1 homepage . Document \\
\end{DL}

syntactic sugar:

\begin{gcotable}
class & Human & - & - & E, F & $\sqcup$ \\
class & E & - & - & = 1 name.string, = 1 givenName.string & $\sqcap$ \\
class & F & - & - & = 1 mbox.IRI, = 1 homepage.Document & $\sqcap$ \\
property & = 1 name.string & name & - & string & = & 1 \\
property & = 1 givenName.string & givenName & - & string & = & 1 \\
property & = 1 mbox.IRI & mbox & - & IRI & = & 1 \\
property & = 1 homepage.Document & homepage & - & Document & = & 1 \\
\end{gcotable}

\section{Allowed Values}

\ms{Allowed Values} corresponds to the requirements:
\ms{R-30-ALLOWED-VALUES-FOR-} \ms{RDF-OBJECTS} and 
\ms{R-37-ALLOWED-VALUES-FOR-RDF-LITERALS}.
It is a common requirement to narrow down the value space of a property by an exhaustive enumeration of the valid values (both literals or resource). This is often rendered in drop down boxes or radio buttons in user interfaces. 
Allowed values for properties
\begin{itemize}
	\item must be these IRIs,
  \item must be IRIs matching specific patterns,
  \item must be IRIs matching one of multiple patterns,
  \item must be (any) literals,
  \item must be literals of a list of allowed literals (e.g. "red" "blue" "green"),
  \item must be typed literals of this type (e.g. XML dataType).
\end{itemize}

\subsection{Example}

\begin{ex}
# DSP:
descriptionTemplate 
    a dsp:DescriptionTemplate ;
    dsp:minOccur "0"^^xsd:nonNegativeInteger ; 
    dsp:maxOccur "infinity"^^xsd:string ; 
    dsp:resourceClass swrc:Book ; 
    dsp:statementTemplate [
        a dsp:NonLiteralStatementTemplate ;
        dsp:minOccur "0"^^xsd:nonNegativeInteger ; 
        dsp:maxOccur "infinity"^^xsd:string ; 
        dsp:property dcterms:subject ; 
        dsp:nonLiteralConstraint [ 
            a dsp:NonLiteralConstraint ;
            dsp:valueClass skos:Concept ;
            dsp:valueURI ComputerScience, SocialScience, Librarianship ] ] .
\end{ex}

\begin{ex}
# OWL2:
dcterms:subject rdfs:range ObjectOneOf . 
# EquivalentClasses( ObjectOneOf ObjectOneOf( ComputerScience SocialScience Librarianship ) )
ObjectOneOf owl:equivalentClass [ 
    a owl:Class ;
    owl:oneOf ( ComputerScience SocialScience Librarianship ) ] .
\end{ex}

The range of the object property \ms{dcterms:subject} must consist of the individuals \ms{ComputerScience} \ms{SocialScience} \ms{Librarianship} which are of the class \ms{skos:Concept}:

\begin{DL}
$\top$ $\sqsubseteq$ $\forall$ dcterms:subject . skos:Concept $\sqcap$ \\
( $\{ComputerScience\}$ $\sqcup$ \{SocialScience\} $\sqcup$ \{Librarianship\} ) \\
\end{DL}

\subsection{Simple Example}

\begin{DL}
Beatle $\equiv$ \{john\} $\sqcup$ \{paul\} $\sqcup$ \{george\} $\sqcup$ \{ringo\} \\
\end{DL}

\begin{gcotable}
class & Beatle & - & - & $\{john\}$, $\{paul\}$, $\{george\}$, $\{ringo\}$ & $\sqcup$ \\
\end{gcotable}

\subsection{Simple Example}

\begin{DL}
\{ComputerScience\} $\sqcup$ \{SocialScience\} $\sqcup$ \{Librarianship\} \\
\end{DL}

\begin{gcotable}
class & complex concept & - & - & \{ComputerScience\}, \{SocialScience\}, \{Librarianship\} & $\sqcup$ & - \\
\end{gcotable}

\subsection{Simple Example}

Jedis have blue, green, or white laser swords.

\begin{ex}
# DSP:
personDescriptionTemplate
    a dsp:DescriptionTemplate ;
    dsp:resourceClass Jedi ;
    dsp:statementTemplate [
        a dsp:LiteralStatementTemplate ;
        dsp:property laserSwordColor ;
        dsp:literalConstraint [
            a dsp:LiteralConstraint ;
            dsp:literal "blue" ;
            dsp:literal "green" ;
            dsp:literal "white"] ] .
\end{ex}

\begin{ex}
# OWL2:
laserSwordColor rdfs:range laserSwordColors . 
laserSwordColors
    a rdfs:Datatype .
    owl:oneOf ( "blue" "green" "white" ) .
\end{ex}

\begin{ex}
# ReSh:
Jedi a rs:ResourceShape ;
    rs:property [
        rs:name "laserSwordColor" ;
        rs:propertyDefinition laserSwordColor ;
        rs:allowedValue "blue" , "green" , "white" ;
        rs:occurs rs:Exactly-one ; ] .
\end{ex}

\begin{ex}
# ShEx:
Jedi {
    laserSwordColor ('blue' 'green' 'white') }
\end{ex}

\begin{ex}
# Jedi individuals:
Yoda 
    laserSwordColor 'blue' .
\end{ex}

\begin{DL}
Jedi $\equiv$ $\exists$ laserSwordColor.\{blue, green, white\} \\
\end{DL}

\begin{gcotable}
property & Jedi & laserSwordColor & - & LaserSwordColor & $\exists$ & - \\
class & LaserSwordColor & - & - & $\{blue\}$, $\{green\}$, $\{white\}$ & $\sqcup$ & - \\
\end{gcotable}

\section{Not Allowed Values}

\ms{Not Allowed Values} corresponds to the requirements
\ms{R-33-NEGATIVE-OBJECT-} \ms{CONSTRAINTS} and
\ms{R-200-NEGATIVE-LITERAL-CONSTRAINTS}.
A matching triple has any literal / object except those explicitly excluded.

\subsection{Simple Example}

Siths do not have blue, green, or white laser swords.

\begin{ex}
# OWL2:
laserSwordColor rdfs:range negativeLaserSwordColors . 
NegativeLaserSwordColors
    a rdfs:Datatype .
    owl:complementOf laserSwordColors .
laserSwordColors
    a rdfs:Datatype .
    owl:oneOf ( "blue" "green" "white" ) .
\end{ex}

\begin{ex}
# ShEx:
Sith {
    ! laserSwordColor ('blue' 'green' 'white') }
\end{ex}

\begin{ex}
# Sith individuals:
DarthSidious
    laserSwordColor 'red' .
\end{ex}

\begin{DL}
Sith $\equiv$ $\neg$ $\exists$ laserSwordColor . \{blue, green, white\} \\
\end{DL}

\begin{gcotable}
property & $\exists$ laserSwordColor . LaserSwordColor & laserSwordColor & - & LaserSwordColor & $\exists$ & - \\
class & Sith & - & - & $\exists$ laserSwordColor . LaserSwordColor & $\neg$ & - \\
class & LaserSwordColor & - & - & $\{blue\}$, $\{green\}$, $\{white\}$ & $\sqcup$ & - \\
\end{gcotable}

\section{Membership in Controlled Vocabularies}

\ms{Membership in Controlled Vocabularies} corresponds to the requirements \\
\ms{R-32-MEMBERSHIP-OF-RDF-OBJECTS-IN-CONTROLLED-VOCABULARIES} and \\
\ms{R-39-MEMBERSHIP-OF-RDF-LITERALS-IN-CONTROLLED-VOCABULARIES}
Resources can only be members of listed controlled vocabularies.

\subsection{Simple Example}

\begin{ex}
# DSP:
bookDescriptionTemplate 
    a dsp:DescriptionTemplate ;
    dsp:resourceClass swrc:Book ; 
    dsp:statementTemplate [
        a dsp:NonLiteralStatementTemplate ;
        dsp:property dcterms:subject ; 
        dsp:nonLiteralConstraint [ 
            a dsp:NonLiteralConstraint ;
            dsp:valueClass skos:Concept ; 
            dsp:vocabularyEncodingScheme BookSubjects, BookTopics, BookCategories ] ] .
\end{ex}

\ms{skos:Concept} resources can only be members of the listed controlled vocabularies.

\begin{DL}
$A \equiv Concept \sqcap B$ \\
$B \equiv \forall inScheme . C$ \\
$C \equiv ConceptScheme \sqcap ( \{BookSubjects\} \sqcup \{BookTopics\} \sqcup \{BookCategories\} )$
\end{DL}

\begin{gcotable}
property & $\top$ & subject & - & A & range & - \\
class & A & - & - & Concept, B & $\sqcap$ & - \\
property & B & inScheme & - & C & $\forall$ & - \\
class & C & - & - & ConceptScheme, D & $\sqcap$ & - \\
class & D & - & - & $\{BookSubjects\}$, $\{BookTopics\}$, $\{BookCategories\}$ & $\sqcup$ & - \\
\end{gcotable}

\begin{ex}
# valid data:
ArtficialIntelligence
    a swrc:Book ;
    dcterms:subject ComputerScience .
ComputerScience
    a skos:Concept ;
    dcam:memberOf BookSubjects ;
    skos:inScheme BookSubjects .
BookSubjects
    a skos:ConceptScheme .
\end{ex}

\begin{ex}
# invalid data:
ArtficialIntelligence
    a swrc:Book ;
    dcterms:subject ComputerScience .
ComputerScience
    a skos:Concept ;
    dcam:memberOf BooksAboutBirds ;
    skos:inScheme BooksAboutBirds ;
    dcam:memberOf BookSubjects ;
    skos:inScheme BookSubjects .
BookSubjects
    a skos:ConceptScheme .
\end{ex}

The related subject (\ms{ComputerScience}) is a member of a controlled vocabulary (\ms{BooksAboutBirds}) 
which is not part of the list of allowed controlled vocabularies.

\section{IRI Pattern Matching}

\ms{IRI Pattern Matching} corresponds to the requirements
\begin{itemize}
	\item \ms{R-21-IRI-PATTERN-MATCHING-ON-RDF-SUBJECTS},
	\item \ms{R-22-IRI-PATTERN-MATCHING-ON-RDF-OBJECTS} and
	\item \ms{R-23-IRI-PATTERN-MATCHING-ON-RDF-PROPERTIES} 
\end{itemize}
indicating IRI pattern matching on subjects, properties, and objects.

\begin{DL}
Not expressible in DL!
\end{DL}

\section{Literal Pattern Matching}

\ms{Literal Pattern Matching} corresponds to the requirement \\
\ms{R-44-PATTERN-MATCHING-ON-RDF-LITERALS}.
indicating pattern matching on literals.

\subsection{Simple Example}

\begin{ex}
# OWL 2 DL (functional-style syntax):
Declaration( Datatype( SSN ) ) 
DatatypeDefinition( 
    SSN
    DatatypeRestriction( xsd:string xsd:pattern "[0-9]{3}-[0-9]{2}-[0-9]{4}" ) )     
DataPropertyRange( hasSSN SSN ) 
\end{ex}

\begin{ex}
# OWL 2 DL (turtle syntax):
SSN 
    a rdfs:Datatype ;
    owl:equivalentClass [
        a rdfs:Datatype ;
        owl:onDatatype xsd:string ;
        owl:withRestrictions ( 
            [ xsd:pattern "[0-9]{3}-[0-9]{2}-[0-9]{4}" ] ) ] .
hasSSN rdfs:range SSN .
\end{ex}

A social security number is a string that matches the given regular expression. 
The second axiom defines SSN as an abbreviation for a datatype restriction on xsd:string. 
The first axiom explicitly declares SSN to be a datatype. 
The datatype SSN can be used just like any other datatype; 
for example, it is used in the third axiom to define the range of the hasSSN property. 

\begin{ex}
# valid data:
TimBernersLee
    hasSSN "123-45-6789"^^SSN .
\end{ex}

\begin{ex}
# invalid data:
TimBernersLee
    hasSSN "123456789"^^SSN .
\end{ex}

\begin{DL}
Not expressible in DL!
\end{DL}

\begin{gcotable}
class & SSN & - & - & xsd:string & xsd:pattern & '[0-9]{3}-[0-9]{2}-[0-9]{4}' \\
\end{gcotable}

\subsection{Simple Example}

There are multiple use cases associated with the requirement to match literals according to given patterns.

Luke's droids can only have the numbers "R2-D2" or "C-3PO".
The universal restriction part of this constraint can be expressed by OWL 2 DL:
\ms{LukesDroids $\sqsubseteq$ $\forall$ droidNumber.DroidNumber}.
The restriction of the datatype \ms{DroidNumber}, however, cannot be expressed in DL, but OWL 2 DL can be used anyway:

\begin{ex}
DroidNumber 
    a rdfs:Datatype ;
    owl:equivalentClass [
        a rdfs:Datatype ;
        owl:onDatatype xsd:string ;
        owl:withRestrictions ( 
            [ xsd:pattern "R2-D2|C-3PO" ] ) ] .
\end{ex}

The second axiom defines \ms{DroidNumber} as an abbreviation for a datatype restriction on \ms{xsd:string}. 
The first axiom explicitly declares \ms{DroidNumber} to be a datatype. 
The datatype \ms{DroidNumber} can be used just like any other datatype like in the universal restriction above.
The literal pattern matching constraint validates \ms{DroidNumber} literals according to the stated regular expression causing a constraint violation for the triples 
\ms{Luke hasDroid Droideka} and \ms{Droideka droidNumber "Droideka"\textasciicircum{}\textasciicircum{}DroidNumber}, 
but not for the triples \ms{Luke hasDroid R2-D2} and \ms{R2-D2 droidNumber "R2-D2"\textasciicircum{}\textasciicircum{}DroidNumber}.

\begin{DL}
Not expressible in DL!
\end{DL}

\begin{gcotable}
property & LukesDroids & droidNumber & - & DroidNumber & $\forall$ & - \\
class & DroidNumber & - & - & xsd:string & xsd:pattern & 'R2-D2$|$C-3PO' \\
\end{gcotable}

\section{Negative Literal Pattern Matching}

\ms{Negative Literal Pattern Matching} corresponds to the requirement \\
\ms{R-44-PATTERN-MATCHING-ON-RDF-LITERALS}
indicating negative pattern matching on literals.

\begin{ex}
# examples:
1. dbo:isbn format is different ’!’ from '^([iIsSbBnN 0-9-])*$'
2. dbo:postCode format is different ‘!’ from 'ˆ[0-9]{5}$'
3. foaf:phone contains any letters ('[A-Za-z]')
\end{ex}

\subsection{Example}

\begin{ex}
# test binding (DQTP):
dbo:isbn format is different ’!’ from '^([iIsSbBnN 0-9-])*$'

P1 => dbo:isbn
NOP => !
REGEX => 'ˆ([iIsSbBnN 0-9-])*$'
\end{ex}

\begin{ex}
# DQTP:
SELECT DISTINCT ?s WHERE { ?s %%P1%% ?value .
    FILTER ( %%NOP%% regex(str(?value), %%REGEX%) ) }
\end{ex}

\begin{itemize}
	\item MATCH Pattern \cite{Kontokostas2014} 
  \item P1 is the property we need to check against REGEX and
NOP can be a not operator (!) or empty.
\end{itemize}

\begin{ex}
# valid data:
FoundationsOfSWTechnologies
    dbo:isbn 'ISBN-13 978-1420090505' .
\end{ex}

\begin{ex}
# invalid data:
HandbookOfSWTechnologies
    dbo:isbn 'DOI 10.1007/978-3-540-92913-0' .
\end{ex}

\begin{DL}
Not expressible in DL!
\end{DL}

\begin{gcotable}
property & $\top$ & dbo:isbn & - & $\neg$ A & domain & - \\
class & $\neg$ A & - & - & A & $\neg$ & - \\
class & A & - & - & xsd:string & regex & 'ˆ([iIsSbBnN 0-9-])*$' $ \\
\end{gcotable}

\section{Literal Value Comparison}

\ms{Literal Value Comparison} corresponds to the requirement \\
\ms{R-43-LITERAL-VALUE-COMPARISON}.
Examples are:

\begin{itemize}
	\item dbo:deathDate before ‘\textless’ dbo:birthDate
  \item dbo:releaseDate after ‘\textgreater’ dbo:latestReleaseDate
  \item dbo:demolitionDate before ‘\textless’ dbo:buildingStartDate
\end{itemize}

\subsection{Simple Example}

\begin{ex}
# DQTP:
SELECT ?s WHERE { 
    ?s %%P1%% ?v1 .
    ?s %%P2%% ?v2 .
    FILTER ( ?v1 %%OP%% ?v2 ) }
\end{ex}

This constraint corresponds to the COMP Pattern \cite{Kontokostas2014}. 
Depending on the property semantics,
there are cases where two different literal values must have
a specific ordering with respect to an operator. 
P1 and P2 are the datatype properties we need to compare and 
OP is the comparison operator (\textless, \textless=, \textgreater, \textgreater=, =, !=). 

\begin{ex}
# test binding (DQTP):
dbo:deathDate before ‘<’ dbo:birthDate

P1 => dbo:deathDate
P2 => dbo:birthDate
OP => <
\end{ex}

\begin{ex}
# valid data:
:AlbertEinstein
    dbo:birthDate '1879-03-14'^^xsd:date ;
    dbo:deathDate '1955-04-18'^^xsd:date .
\end{ex}

\begin{ex}
# invalid data:
:NeilArmstrong
    dbo:birthDate '2012-08-25'^^xsd:date ;
    dbo:deathDate '1930-08-05'^^xsd:date .
\end{ex}

\begin{DL}
Not expressible in DL!
\end{DL}

\begin{gcotable}
property & $\top$ & dbo:birthDate & dbo:deathDate & xsd:date & \textgreater & - \\
\end{gcotable}

\subsection{Simple Example}

Duplicate strings are not allowed:

\begin{ex}
dc:subject "foo"@en
dc:subject "foo"@fr
dc:subject "bar"@fr
dc:subject "foo"
\end{ex}

There are no identical strings.

\begin{DL}
Not expressible in DL!
\end{DL}

\begin{gcotable}
property & $\top$ & dc:subject & dc:subject & xsd:string & $\neq$ & - \\
\end{gcotable}

\section{Negative Literal Ranges}

\ms{Negative Literal Ranges} corresponds to the requirement
\ms{R-142-NEGATIVE-} \ms{RANGES-OF-RDF-LITERAL-VALUES}.
The literal value of a resource (having a certain type) must (not) be within a specific range.
P1 is a data property of an instance of class T1 and its literal value must be between the range of [Vmin,Vmax] or outside ( ‘!' ).

\subsection{Simple Example}

\begin{itemize}
	\item dbo:height of a dbo:Person is not within [0.4,2.5]
\end{itemize}

\subsection{Simple Example}

\begin{itemize}
	\item geo:lat of a spatial:Feature is not within [-90,90]
\end{itemize}

\subsection{Simple Example}

\begin{itemize}
	\item geo:long of a gml:Feature must be in range [-180,180]
\end{itemize}

\section{Literal Ranges}

\ms{Literal Ranges} corresponds to the requirement
\ms{R-45-RANGES-OF-RDF-LITERAL-} \ms{VALUES}.
P1 is a data property (of an instance of class C1) and its literal value must be between the range of [$V_{min}$,$V_{max}$].

\subsection{Example}

\begin{ex}
# OWL 2 DL (functional-style syntax):
Declaration( Datatype( NumberPlayersPerWorldCupTeam ) ) 
DatatypeDefinition( 
    NumberPlayersPerWorldCupTeam
    DatatypeRestriction( 
        xsd:nonNegativeInteger 
        xsd:minInclusive "1"^^xsd:nonNegativeInteger 
        xsd:maxInclusive "23"^^xsd:nonNegativeInteger ) )     
DataPropertyRange( position NumberPlayersPerWorldCupTeam ) 
\end{ex}

\begin{ex}
# OWL 2 DL (turtle syntax):
NumberPlayersPerWorldCupTeam
    a rdfs:Datatype ;
    owl:equivalentClass [
        a rdfs:Datatype ;
        owl:onDatatype xsd:nonNegativeInteger ;
        owl:withRestrictions ( 
            [ xsd:minInclusive "1"^^xsd:nonNegativeInteger ]
            [ xsd:maxInclusive "23"^^xsd:nonNegativeInteger ] ) ] .
position rdfs:range NumberPlayersPerWorldCupTeam .
\end{ex}

The data range 'NumberPlayersPerWorldCupTeam' contains the non negative integers 1 to 23, as each world cup team can only have 23 football players at most.

\begin{ex}
# valid data:
MarioGoetze
    position "19"^^:NumberPlayersPerWorldCupTeam .
\end{ex}

\begin{ex}
# invalid data:
MarioGoetze
    position "99"^^:NumberPlayersPerWorldCupTeam .
\end{ex}



\begin{DL}
Not expressible in DL!
\end{DL}

%

\begin{gcotable}
class & NumberPlayersPerWorldCupTeam & - & - & $\geq$1, $\leq$23 & $\sqcap$ & - \\
class & $\geq$1 & - & - & xsd:nonNegativeInteger & $\geq$ & 1 \\
class & $\leq$23 & - & - & xsd:nonNegativeInteger & $\leq$ & 23 \\
\end{gcotable}

%

\section{Ordering}

\ms{Ordering} corresponds to the requirements
\ms{R-121-SPECIFY-ORDER-OF-RDF-} \ms{RESOURCES} and
\ms{R-217-DEFINE-ORDER-FOR-FORMS/DISPLAY}.
With this constraint objects of object properties can be ordered as well as literals of data properties and objects of object properties.

\subsection{Simple Example}

Define the order of the property \ms{listElement} for the class \ms{rdf:List}.
List elements are of datatype \ms{xsd:string}.

\begin{DL}
Not expressible in DL!
\end{DL}

\begin{gcotable}
property & rdf:List & listElement & - & xsd:string & order & 1 \\
\end{gcotable}

\subsection{Simple Example}

Property \emph{p} should be ordered, i.e., point to a list (rdf:List) of values.

\begin{gcotable}
property & $\top$ & p & - & - & ordered property & - \\
\end{gcotable}

\section{Validation Levels}

\ms{Validation Levels} corresponds to the requirements

\begin{itemize}
	\item \ms{R-205-VARYING-LEVELS-OF-ERROR},
	\item \ms{R-135-CONSTRAINT-LEVELS},
	\item \ms{R-158-SEVERITY-LEVELS-OF-CONSTRAINT-VIOLATIONS}, and
	\item \ms{R-193-MULTIPLE-CONSTRAINT-VALIDATION-EXECUTION-LEVELS}.
\end{itemize}

Different levels of severity, priority should be assigned to constraints.
Possible validation levels could be: informational, warning, error, fail, should, recommended, must, may, optional, closed (only this) constraints, open (at least this).

\begin{DL}
Not expressible in DL!
\end{DL}

\section{String Operations}

\ms{String Operations} corresponds to the requirement
\ms{R-194-PROVIDE-STRING-} \ms{FUNCTIONS-FOR-RDF-LITERALS}.
Some constraints require building new strings out of other strings.
Calculating the string length would also be another constraint of this type.

\subsection{Simple Example}

The length of strings of the property \ms{hasISBN} for the context class \ms{Book} must be exactly 3.

\begin{DL}
Not expressible in DL!
\end{DL}

\begin{gcotable}
property & Book & hasISBN & - & xsd:string & length & 3 \\
\end{gcotable}

\section{Context-Specific Valid Classes}

\ms{Context-Specific Valid Classes} corresponds to the requirements \\
\ms{R-209-VALID-CLASSES}.
What types of resources (rdf:type) are valid in a specific context? 
Context can be an input stream, a data creation function, or an API.

\subsection{Simple Example}

Within the context \ms{AP} (application profile) the classes \ms{Book} and \ms{Paper} are valid.

\begin{DL}
Not expressible in DL!
\end{DL}

\begin{gcotable}
class & AP & - & - & Book, Paper & context-specific valid classes & true \\
\end{gcotable}

\section{Context-Specific Valid Properties}

\ms{Context-Specific Valid Properties} corresponds to the requirement
\ms{R-210-} \ms{VALID-PROPERTIES}.
What properties can be used within this context? 
Context can be an data receipt function, data creation function, or API.

\subsection{Simple Example}

Within the context \ms{AP} (application profile) the properties \ms{dcterms:subject} and \ms{dcterms:title} are invalid.

\begin{DL}
Not expressible in DL!
\end{DL}

\begin{gcotable}
property & AP & dcterms:subject, dcterms:title & - & - & context-specific valid properties & false \\
\end{gcotable}

\section{Default Values}

\ms{Default Values} corresponds to the requirements
\ms{R-31-DEFAULT-VALUES-OF-RDF-} \ms{OBJECTS} and
\ms{R-38-DEFAULT-VALUES-OF-RDF-LITERALS}.
Default values for objects and literals are inferred automatically.
It should be possible to declare the default value for a given property, e.g. so that input forms can be pre-populated and to insert a required property that is missing in a web service call. 

\subsection{Simple Example}

Jedis have only 1 blue laser sword per default.
Siths, in contrast, normally have 2 red laser swords.

\begin{ex}
# rule (SPIN)
# -----------
owl:Thing
    spin:rule [
        a sp:Construct ;
            sp:text """
                CONSTRUCT {            
                    ?this laserSwordColor "blue"^^xsd:string ;
                    ?this numberLaserSwords "1"^^xsd:nonNegativeInteger . 
                }
                WHERE {             
                    ?this a Jedi .            
                } """ ; ] .
owl:Thing
    spin:rule [
        a sp:Construct ;
            sp:text """
                CONSTRUCT {
                    ?this laserSwordColor "red"^^xsd:string ;
                    ?this numberLaserSwords "2"^^xsd:nonNegativeInteger . 
                }
                WHERE {             
                    ?this a Sith .            
                } """ ; ] .
\end{ex}

\begin{ex}
# data:
Joda a Jedi .
DarthSidious a Sith .
\end{ex}

\begin{ex}
# inferred triples:
Joda 
    laserSwordColor "blue"^^xsd:string ;
    numberLaserSwords "1"^^xsd:nonNegativeInteger .
DarthSidious 
    laserSwordColor "red"^^xsd:string ;
    numberLaserSwords "2"^^xsd:nonNegativeInteger .
\end{ex}

\begin{DL}
Not expresible in DL!
\end{DL}

\begin{gcotable}
property & Jedi & laserSwordColor & - & - & default values & 'blue' \\
property & Jedi & numberLaserSwords & - & - & default values & '1' \\
property & Sith & laserSwordColor & - & - & default values & 'red' \\
property & Sith & numberLaserSwords & - & - & default values & '2' \\
\end{gcotable}

\section{Mathematical Operations}

\ms{Mathematical Operations} corresponds to the requirements
\begin{itemize}
	\item \ms{R-42-MATHEMATICAL-OPERATIONS} and
  \item \ms{R-41-STATISTICAL-COMPUTATIONS}.
\end{itemize}
Examples are:
\begin{itemize}
	\item adding 2 dates
	\item add number of days to start date
	\item area = width * height
	\item Statistical Computations: average, mean, sum
\end{itemize}

\subsection{Simple Example}

Calculate rectangle areas.

\begin{DL}
Not expressible in DL!
\end{DL}

\begin{gcotable}
property & Rectangle & area & width, height & xsd:integer & multiplication & - \\
\end{gcotable}

\section{Language Tag Matching}

\ms{Language Tag Matching} corresponds to the requirement \\
\ms{R-47-LANGUAGE-TAG-MATCHING}.

\subsection{Simple Example}

Only English names are allowed.

\begin{DL}
Not expressible in DL!
\end{DL}

\begin{gcotable}
property & $\top$ & name & - & - & language tag & 'en' \\
\end{gcotable}

\section{Language Tag Cardinality}

\ms{Language Tag Cardinality} corresponds to the requirements
\begin{itemize}
	\item \ms{R-49-RDF-LITERALS-HAVING-AT-MOST-ONE-LANGUAGE-TAG} and
	\item \ms{R-48-MISSING-LANGUAGE-TAGS}.
\end{itemize}

\subsection{Simple Example}

Check that no language is used more than once per property

\begin{ex}
# DQTP:
SELECT DISTINCT ?s WHERE { ?s %%P1%% ?c
    BIND ( lang(?c) AS ?l )
    FILTER (isLiteral (?c) && lang(?c) = %%V1%%)}
GROUP BY ?s HAVING COUNT (?l) > 1
\end{ex}

This corresponds to the test pattern \ms{ONELANGPattern} \cite{Kontokostas2014}.
A literal value should contain at most 1 literal for a language. 
\ms{P1} is the property containing the literal and \ms{V1} is the language we want to check.

\begin{ex}
# test binding (DQTP):
P1 => foaf:name
V1 => en
\end{ex}

A single English (“en”) \ms{foaf:name}.

\begin{ex}
# valid data:
:LeiaSkywalker
    foaf:name 'Leia Skywalker'@en .
\end{ex}

\begin{ex}
# invalid data:
:LeiaSkywalker
    foaf:name 'Leia Skywalker'@en ;
    foaf:name 'Leia'@en .
\end{ex}

\begin{DL}
Not expressible in DL!
\end{DL}

\begin{gcotable}
property & $\top$ & foaf:name & - & - & language tag exact cardinality & 1 \\
\end{gcotable}

\section{Whitespace Handling}

\ms{Whitespace Handling} corresponds to the requirement
\ms{R-50-WHITESPACE-} \\
\ms{HANDLING-OF-RDF-LITERALS}.
Avoid whitespaces in literals neither leading nor trailing white spaces.

\subsection{Simple Example}

Check if literals of the property \ms{p} do not include whitespaces.

\begin{DL}
Not expressible in DL!
\end{DL}

\begin{gcotable}
property & $\top$ & p & - & - & whitespace & - \\
\end{gcotable}

\section{HTML Handling}

\ms{HTML Handling} corresponds to the requirement
\ms{R-51-HTML-HANDLING-OF-RDF-} \ms{LITERALS}.
Check if there are no HTML tags included in literals (of specific data properties within the context of specific classes).

\subsection{Simple Example}

Check if literals of the property \ms{p} of the class \ms{C} do not include HTML tags.

\begin{DL}
Not expressible in DL!
\end{DL}

\begin{gcotable}
property & c & p & - & - & HTML & - \\
\end{gcotable}

\section{Required Properties}

\ms{Required Properties} corresponds to the requirement \ms{R-68-REQUIRED-PROPERTIES}.

\subsection{Simple Example}

For persons the property \ms{hasAncestor} has to be stated pointing to persons.

\begin{DL}
$Person \sqsubseteq \exists hasAncestor.Person$
\end{DL}

\begin{gcotable}
property & Person & hasAncestor & - & Person & $\exists$ & - \\
\end{gcotable}

\section{Optional Properties}

\ms{Optional Properties} corresponds to the requirement \ms{R-69-OPTIONAL-PROPERTIES}.

\subsection{Simple Example}

For persons the property \ms{hasAncestor} pointing to persons is optional.

\begin{DL}
$\exists hasAncestor.Person \sqsubseteq Person$ \\
\end{DL}

Same as the definition of domains.

\begin{gcotable}
property & $\exists$ hasAncestor.Person & hasAncestor & - & Person & $\exists$ & - \\
class & Person & - & - & $\exists$ hasAncestor.Person & $\sqsubseteq$ & - \\
\end{gcotable}	

syntactic sugar:

\begin{gcotable}
property & Person & hasAncestor & - & Person & optional & - \\
\end{gcotable}

\section{Repeatable Properties}

\ms{Repeatable Properties} corresponds to the requirement \\
\ms{R-70-REPEATABLE-PROPERTIES}.

\subsection{Simple Example}

The property \ms{commandsVessel} is repeatable for individuals of the class \ms{Captain}.

\begin{DL}
$Captain \sqsubseteq \geq1 commandsVessel.\top $
\end{DL}

\begin{gcotable}
property & Captain & commandsVessel & - & $\top$ & $\geq$ & 1 \\
\end{gcotable}

syntactic sugar:

\begin{gcotable}
property & Captain & commandsVessel & - & $\top$ & repeatable & - \\
\end{gcotable}

\section{Conditional Properties}

\ms{Conditional Properties} corresponds to the requirement \\
\ms{R-71-CONDITIONAL-PROPERTIES}.

Multiple conditions are possible:

\begin{itemize}
  \item universal quantification on object and data properties,
  \item existential quantification on object and data properties,
  \item if specific properties are present, then specific other properties also have to be present
\end{itemize}

\subsection{Simple Example}

If an individual has a \ms{parentOf} property relationship, then this individual also has a \ms{ancestorOf} property relationship.

\begin{DL}
parentOf $\sqsubseteq$ ancestorOf 
\end{DL}

\begin{gcotable}
property & $\top$ & parentOf & ancestorOf & $\top$ & $\sqsubseteq$ \\
\end{gcotable}

\section{Recommended Properties}

\ms{Recommended Properties} corresponds to the requirement \\
\ms{R-72-RECOMMENDED-PROPERTIES}.
Which properties are not required but recommended within a particular context.

\subsection{Simple Example}

Property \ms{p} is recommended to use within the context of the class \ms{C}.

\begin{DL}
Not expressible in DL!
\end{DL}

\begin{gcotable}
property & C & p & - & - & recommended & - \\
\end{gcotable}

\section{Negative Property Constraints}

\ms{Negative Property Constraints} corresponds to the requirements
\begin{itemize}
	\item \ms{R-52-NEGATIVE-OBJECT-PROPERTY-CONSTRAINTS} and
	\item \ms{R-53-NEGATIVE-DATA-PROPERTY-CONSTRAINTS}.
\end{itemize}
Instances of a specific class must not have some object property.
In OWL 2 DL, this can be expressed as follows: \ms{ObjectComplementOf ( ObjectSomeValuesFrom ( ObjectPropertyExpression owl:Thing ) )}.

\subsection{Example}

\begin{ex}
# ShEx:
<FeelingForce> {
    feelingForce (true) ,
    attitute xsd:string }
<JediMentor> {
    feelingForce (true) ,
    attitute ('good') ,
    laserSwordColor xsd:string ,
    numberLaserSwords xsd:nonNegativeInteger ,
    mentorOf @<JediStudent> ,
   !studentOf @<JediMentor> }
<JediStudent> {
    feelingForce (true) ,
    attitute ('good') ,
    laserSwordColor xsd:string ,
    numberLaserSwords xsd:nonNegativeInteger ,
   !mentorOf @<JediStudent> ,
    studentOf @<JediMentor> }
\end{ex}

A matching triple has any predicate except those excluded by the '!' operator.

\begin{ex}
# individuals matching 'FeelingForce' and 'JediMentor' data shapes:
Obi-Wan 
    feelingForce true ;
    attitute 'good' ;
    laserSwordColor 'blue' ;
    numberLaserSwords '1'^^xsd:nonNegativeInteger ;
    mentorOf Anakin .
\end{ex}

\begin{ex}
# individuals matching 'FeelingForce' and 'JediStudent' data shapes:
Anakin 
    feelingForce true ;
    attitute 'good' ;
    laserSwordColor 'blue' ;
    numberLaserSwords '1'^^xsd:nonNegativeInteger ;
    studentOf Obi-Wan .
\end{ex}

\begin{DL}
$JediMentor \sqsubseteq \neg(\exists studentOf.JediMentor)$
\end{DL}

\begin{gcotable}
property & $\exists$ studentOf.JediMentor & studentOf & - & JediMentor & $\exists$ & - \\
class & JediMentor & - & - & $\exists$ studentOf.JediMentor & $\neg$ & - \\
\end{gcotable}

\section{Handle RDF Collections}

\ms{Handle RDF Collections} corresponds to the requirement
\ms{R-120-HANDLE-RDF-} \ms{COLLECTIONS}.
Examples are:
\begin{itemize}
	\item size of collection
	\item first / last element of list must be a specific literal
	\item compare elements of collection
	\item are collections identical?
	\item actions on RDF lists\footnote{See \url{http://www.snee.com/bobdc.blog/2014/04/rdf-lists-and-sparql.html}}
	\item 2. list element equals 'XXX'
	\item Does the list have more than 10 elements?
\end{itemize}

\subsection{Example}

Get 2. list element:

\begin{ex}
# SPIN:
getListItem
    a spin:Function ; rdfs:subClassOf spin:Functions ;
    spin:constraint [
        rdf:type spl:Argument ;
        spl:predicate sp:arg1 ;
        spl:valueType rdf:List ;
        rdfs:comment "list" ; ] ;
    spin:constraint [
        rdf:type spl:Argument ;
        spl:predicate sp:arg2 ;
        spl:valueType xsd:nonNegativeInteger ;
        rdfs:comment "item position (starting with 0)" ; ] ;
    spin:body [
        a sp:SELECT ;
        sp:text """
            SELECT ?item
            WHERE {
                ?arg1 contents/rdf:rest{?arg2}/rdf:first ?item } """ ; ] ;
    spin:returnType rdfs:Resource .
\end{ex}

\begin{ex}
# data:
Jinn students 
     ( Xanatos Kenobi ) . 
\end{ex}

\begin{ex}
# SPIN:
BIND ( getListItem( ?list, "1"xsd:nonNegativeInteger ) AS ?listItem ) .
\end{ex}

\begin{itemize}
  \item SPIN function call
	\item retrieves the 2. item from the list (2. student of Jedi mentor Jinn)
\end{itemize}

\begin{ex}
# result:
Kenobi
\end{ex}

\begin{DL}
Not expressible in DL!
\end{DL}

\subsection{Simple Example}

Append a string list element 'list element' to a list the property \ms{p} of the class \ms{C} is pointing to.

\begin{DL}
Not expressible in DL!
\end{DL}

\begin{gcotable}
property & C & p & - & xsd:string & append list element & 'list element' \\
\end{gcotable}

\section{Recursive Queries}

\ms{Recursive Queries} corresponds to the requirement
\ms{R-222-RECURSIVE-QUERIES}.
If we want to define Resource Shapes, remember that it is a recursive
language (the valueShape of a Resource Shape is in turn another
Resource Shape). There is no way to express that in SPARQL without
hand-waving "and then you call the function again here" or "and then
you embed this operation here" text.  The embedding trick doesn't work
in the general case because SPARQL can't express recursive queries,
e.g. "test that this Issue is valid and all of the Issues that
references, recursively".
Most SPARQL engines already have
functions that go beyond the official SPARQL 1.1 spec. The cost of that
sounds manageable.

\subsection{Simple Example}

\begin{ex}
# ShEx:
IssueShape {
    related @IssueShape*
}
\end{ex}

\begin{DL}
$IssueShape \sqsubseteq \geq1 related.IssueShape $
\end{DL}

\begin{gcotable}
property & IssueShape & related & - & IssueShape & $\geq$ & 1 \\
\end{gcotable}

\section{Value is Valid for Datatype}

Make sure that a value is valid for its datatype.

\subsection{Simple Example}

A date is really a date, as an example.
SPARQL regex can be used for this purpose.

\subsection{Simple Example}

\begin{ex}
# SPIN:
FILTER ( datatype( ?shoeSize ) = xsd:nonNegativeInteger )
isNumeric ( ?shoeSize )
\end{ex}

The datatype of ?showSize is xsd:nonNegativeInteger.
The datatype is really numeric.

\begin{gcotable}
property & $\top$ & shoeSize & - & xsd:nonNegativeInteger & value valid for datatype & - \\
\end{gcotable}

%
%

\section{Individual Equality}

\ms{Individual equality} states that two different names are known to refer to the same individual \cite{Kroetzsch2012}.

\subsection{Simple Example}

\begin{DL}
$\{julia\} = \{john\} $\\
\end{DL}

\begin{gcotable}
class & \{julia\} & - & - & \{john\} & = & - \\
class & \{john\} & - & - & \{julia\} & = & - \\
\end{gcotable}

\section{Individual Inequality}

This is by default because of the UNA.

\subsection{Simple Example}

\begin{DL}
$\{julia\} \ne \{john\} $\\
\end{DL}

Asserts that Julia and John are actually different individuals.

\begin{gcotable}
class & \{julia\} & - & - & \{john\} & $\ne$ & - \\
class & \{john\} & - & - & \{julia\} & $\ne$ & - \\
\end{gcotable}

\section{Equivalent Properties}
An equivalent object properties axiom EquivalentObjectProperties( OPE1 ... OPEn ) states that all of the object property expressions OPEi, 1 $\leq$ i $\leq$ n, are semantically equivalent to each other. This axiom allows one to use each OPEi as a synonym for each OPEj — that is, in any expression in the ontology containing such an axiom, OPEi can be replaced with OPEj without affecting the meaning of the ontology. The axiom EquivalentObjectProperties( OPE1 OPE2 ) is equivalent to the following two axioms SubObjectPropertyOf( OPE1 OPE2 ) and SubObjectPropertyOf( OPE2 OPE1 ).

An equivalent data properties axiom EquivalentDataProperties( DPE1 ... DPEn ) states that all the data property expressions DPEi, 1 $\leq$ i $\leq$ n, are semantically equivalent to each other. This axiom allows one to use each DPEi as a synonym for each DPEj — that is, in any expression in the ontology containing such an axiom, DPEi can be replaced with DPEj without affecting the meaning of the ontology. The axiom EquivalentDataProperties( DPE1 DPE2 ) can be seen as a syntactic shortcut for the following axiom SubDataPropertyOf( DPE1 DPE2 ) and SubDataPropertyOf( DPE2 DPE1 ).

\subsection{Simple Example}

\begin{ex}
# OWL 2:
hasBrother owl:equivalentProperty hasMaleSibling . 
Chris hasBrother Stewie . 
Stewie hasMaleSibling Chris .
\end{ex}

entailments:

\begin{ex}
Chris hasMaleSibling Stewie . 
Stewie hasBrother Chris .
\end{ex}

\begin{DL}
hasBrother $\sqsubseteq$ hasMaleSibling $\sqcap$ hasMaleSibling $\sqsubseteq$ hasBrother \\
\end{DL}

\begin{gcotable}
property & $\top$ & hasBrother & hasMaleSibling & - & $\sqsubseteq$ \\
property & $\top$ & hasMaleSibling & has Brother & - & $\sqsubseteq$ \\
\end{gcotable}

\begin{DL}
$hasBrother \equiv hasMaleSibling$
\end{DL}

\section{Property Assertions}

\ms{{Property Assertions}} corresponds to the requirement \ms{R-96-PROPERTY-ASSERTIONS}
and includes positive property assertions and negative property assertions.
A positive object property assertion ObjectPropertyAssertion( OPE a1 a2 ) states that the individual a1 is connected by the object property expression OPE to the individual a2. 
A negative object property assertion NegativeObjectPropertyAssertion( OPE a1 a2 ) states that the individual a1 is not connected by the object property expression OPE to the individual a2. 
A positive data property assertion DataPropertyAssertion( DPE a lt ) states that the individual a is connected by the data property expression DPE to the literal lt. 
A negative data property assertion NegativeDataPropertyAssertion( DPE a lt ) states that the individual a is not connected by the data property expression DPE to the literal lt.

\subsection{Simple Example}

\begin{ex}
# OWL 2:
NegativeObjectPropertyAssertion( hasSon Peter Meg )
\end{ex}

Meg is not a son of Peter.

\begin{DL}
$hasSon(Peter,Meg)$\\
\end{DL}

The negation of such an assertion is not necessary, as it's meaningless!

\begin{gcotable}
property & \{Peter\} & hasSon & - & \{Meg\} & $\neq$ & - \\
\end{gcotable}

\subsection{Simple Example}

\begin{ex}
# OWL 2:
DataPropertyAssertion( :hasAge :Meg "17"^^xsd:integer )
\end{ex}

Meg is seventeen years old. 

\begin{DL}
$hasAge ( Meg ,"17"\hat{\ } \hat{\ } xsd:integer ) $\\
\end{DL}

\begin{gcotable}
property & \{Meg\} & hasAge & - & - & = & "17"$\hat{\ } \hat{\ }$ xsd:integer \\
\end{gcotable}

\section{Functional Properties}

An object property functionality axiom FunctionalObjectProperty( OPE ) states that the object property expression OPE is functional — that is, for each individual x, there can be at most one distinct individual y such that x is connected by OPE to y. Each such axiom can be seen as a syntactic shortcut for the following axiom: SubClassOf( owl:Thing ObjectMaxCardinality( 1 OPE ) ).

\subsection{Simple Example}

\begin{ex}
# OWL 2:
hasFather rdf:type owl:FunctionalProperty . 	
Stewie hasFather Peter . 	
Stewie hasFather Peter_Griffin . 
\end{ex}

Each object can have at most one father. 

entailment:

\begin{ex}
Peter_Griffin owl:sameAs Peter .  
\end{ex}

\begin{DL}
$(\textsf{funct } \textsf{hasFather})$
\end{DL}

\begin{gcotable}
property & $\top$ & hasFather & - & - & functional & - \\
\end{gcotable}

\section{Inverse-Functional Properties}

An object property inverse functionality axiom InverseFunctionalObjectProperty( OPE ) states that the object property expression OPE is inverse-functional - that is, for each individual x, there can be at most one individual y such that y is connected by OPE with x. Each such axiom can be seen as a syntactic shortcut for the following axiom: SubClassOf( owl:Thing ObjectMaxCardinality( 1 ObjectInverseOf( OPE ) ) ).

\subsection{Simple Example}

\begin{ex}
# OWL 2:
fatherOf rdf:type owl:InverseFunctionalProperty . 	
Peter fatherOf Stewie .
Peter_Griffin fatherOf Stewie .
\end{ex}

Each object can have at most one father. 

Entailment:

\begin{ex}
Peter owl:sameAs Peter_Griffin . 
\end{ex}

\begin{DL}
$(\textsf{funct } \textsf{hasFather}\sp{\overline{\ }})$
\end{DL}

\begin{gcotable}
property & $\top$ & hasFather$^{-}$ & hasFather & - & inverse & - \\
property & $\top$ & hasFather$^{-}$ & - & - & functional & - \\
\end{gcotable}

\section{Value Restrictions}

Individual Value Restrictions: A has-value class expression ObjectHasValue( OPE a ) consists of an object property expression OPE and an individual a, and it contains all those individuals that are connected by OPE to a. Each such class expression can be seen as a syntactic shortcut for the class expression ObjectSomeValuesFrom( OPE ObjectOneOf( a ) ). 
Literal Value Restrictions: A has-value class expression DataHasValue( DPE lt ) consists of a data property expression DPE and a literal lt, and it contains all those individuals that are connected by DPE to lt. Each such class expression can be seen as a syntactic shortcut for the class expression DataSomeValuesFrom( DPE DataOneOf( lt ) ).

\subsection{Simple Example}

\begin{DL}
$FatherOfStewie \sqsubseteq \exists fatherOf.\{Stewie\}$
\end{DL}

\begin{ex}
# OWL 2:
Peter fatherOf Stewie . 
ObjectHasValue( fatherOf Stewie )
\end{ex}

The has-value class expression contains those individuals that are connected through the \ms{fatherOf} property with the individual \ms{Stewie}.
\ms{Peter} is classified as its instance.

\begin{gcotable}
property & FatherOfStewie & fatherOf & - & $\{Stewie\}$ & $\exists$ \\
\end{gcotable}

\section{Self Restrictions}

A self-restriction ObjectHasSelf( OPE ) consists of an object property expression OPE, and it contains all those individuals that are connected by OPE to themselves. 

\subsection{Simple Example}

\begin{ex}
# OWL 2:
Peter likes Peter . 
ObjectHasSelf( likes ) 
\end{ex}

\begin{DL}
LikesThemselves $\sqsubseteq$ $\exists$ likes.Self
\end{DL}

\begin{gcotable}
property & $\exists$ likes . Self & likes & - & Self & $\exists$ \\
class & LikesThemselves & - & - & $\top$, $\exists$ likes . Self & $\sqsubseteq$ \\
\end{gcotable}

The self-restriction contains those individuals that like themselves.
\ms{Peter} is classified as its instance.

\section{Data Property Facets}

\ms{Data Property Facets} corresponds to the requirement
\ms{R-46-CONSTRAINING-FACETS}.
For datatype properties it should be possible to declare frequently needed "facets" to drive user interfaces and validate input against simple conditions, including min/max value, regular expressions, string length etc. similar to XSD datatypes. 
Constraining facets to restrict datatypes of RDF literals.
Constraining facets may be: xsd:length, xsd:minLength, xsd:maxLength, xsd:pattern, xsd:enumeration, xsd:whiteSpace, xsd:maxInclusive, xsd:maxExclusive, xsd:minExclusive, xsd:minInclusive, xsd:totalDigits, xsd:fractionDigits.

\subsection{Simple Example}

\textbf{string matches regular expression}

\begin{ex}
# OWL 2 QL (functional-style syntax):
Declaration( Datatype( SSN ) ) 
DatatypeDefinition( 
    SSN
    DatatypeRestriction( xsd:string xsd:pattern "[0-9]{3}-[0-9]{2}-[0-9]{4}" ) )     
DataPropertyRange( hasSSN SSN ) 
\end{ex}

\begin{ex}
# OWL 2 QL (turtle syntax):
SSN 
    a rdfs:Datatype ;
    owl:equivalentClass [
        a rdfs:Datatype ;
        owl:onDatatype xsd:string ;
        owl:withRestrictions ( 
            [ xsd:pattern "[0-9]{3}-[0-9]{2}-[0-9]{4}" ] ) ] .
hasSSN rdfs:range SSN .
\end{ex}

A social security number is a string that matches the given regular expression. 
The second axiom defines SSN as an abbreviation for a datatype restriction on xsd:string. 
The first axiom explicitly declares SSN to be a datatype. 
The datatype SSN can be used just like any other datatype; 
for example, it is used in the third axiom to define the range of the hasSSN property. 

\begin{ex}
# valid data:
TimBernersLee
    hasSSN "123-45-6789"^^SSN .
\end{ex}

\begin{ex}
# invalid data:
TimBernersLee
    hasSSN "123456789"^^SSN .
\end{ex}

\begin{DL}
Not expressible in DL!
\end{DL}

\begin{gcotable}
class & SSN & - & - & xsd:string & xsd:pattern & '[0-9]{3}-[0-9]{2}-[0-9]{4}' \\
\end{gcotable}

\section{Primary Key Properties}

It is often useful to declare a given (datatype) property as the "primary key" of a class, so that the system can enforce uniqueness and also automatically build URIs from user input and data imported from relational databases or spreadsheets.

\subsection{Simple Example}

The \ms{Primary Key Properties} constraint is often useful to declare a given (datatype) property as the "primary key" of a class, so that a system can enforce uniqueness. 
Starfleet officers, e.g., are uniquely identified by their command authorization code (e.g. to activate and cancel auto-destruct sequences).
It means that the property \ms{commandAuthorizationCode} is inverse functional - mapped to DL and the RDF-CO as follows:

\begin{DL}
$(\ms{funct } commandAuthorizationCode\sp{\overline{\ }})$
\end{DL}

\begin{gcotable}
property & $\top$ & commandAuthorizationCode$^{-}$ & commandAuthorizationCode$^{-}$ & - & inverse & - \\
property & $\top$ & commandAuthorizationCode$^{-}$ & - & - & functional & - \\
\end{gcotable}

Keys, however, are even more general, i.e., a generalization of inverse functional properties \cite{Schneider2009}.
A key can be a datatype property, an object property, or a chain of properties.
For this generalization purposes, as there are different sorts of key, and as keys can lead to undecidability, 
DL is extended with \ms{key boxes} and a special \ms{keyfor} construct\cite{Lutz2005}.
This leads to the following DL and RDF-CO mappings (only one simple property constraint):

\begin{DL}
commandAuthorizationCode \ms{keyfor} StarfleetOfficer
\end{DL}

\begin{gcotable}
property & StarFleetOfficer & commandAuthorizationCode & - & - & keyfor & - \\
\end{gcotable}

\section{Use Sub-Super Relations in Validation}

\ms{Exploiting Class/Property Specialization Ontology Axioms} corresponds to the requirement
\ms{R-224-instance-level-data-validation-exploitingclass/} \ms{property-specialization-axioms-in-ontologies}.
Validation of instances data (direct or indirect) exploiting the subclass or sub-property link in a given ontology.
This validation can indicate when the data is verbose (redundant) or expressed at a too general level, and could be improved.
Examples are:
\begin{itemize}
	\item If dc:date and one of its sub-properties dcterms:created or dcterms:issued are present, check that the value in dc:date is not redundant with dcterms:created or dcterms:issued for ingestion
  \item Check if dc:rights has the same value than edm:rights either as rdf:resource or literal, if yes dc:rights is redundant
  \item If one or more dc:coverage are present, suggest the use of one of its sub-properties, dcterms spatial or dcterms:temporal.
\end{itemize}

\subsection{Simple Example}

If dc:date and one of its sub-properties dcterms:created or dcterms:issued are present, check that the value in dc:date is not redundant with dcterms:created or dcterms:issued for ingestion

\begin{DL}
Not expressible in DL!
\end{DL}

\begin{gcotable}
property & $\top$ & dc:date & dcterms:created, dcterms:issued & - & not redundant & - \\
\end{gcotable}

\section{Cardinality Shortcuts}

In most Library applications, cardinality shortcuts tend to appear in pairs, with repeatable / non-repeatable establishing maximum cardinality and optional / mandatory establishing minimum cardinality.
\begin{itemize}
	\item Optional \& Non-Repeatable = [0,1]
  \item Optional \& Repeatable = [0,*]
  \item Mandatory \& Non-Repeatable = [1,1]
  \item Mandatory \& Repatable = [1,*]
\end{itemize}
Can be expressed in DL using minimum and maximum cardinality restrictions.
We propose to use syntactic sugar instead:
\begin{gcotable}
property & C & p & - & - & optional and non-repeatable & - \\
\end{gcotable}

\section{Aggregations}

Some constraints require aggregating multiple values, especially via COUNT, MIN and MAX. 

\subsection{Simple Example}

\begin{DL}
p = COUNT (q)
\end{DL}

Context class is \ms{C}.

\begin{DL}
Not expressible in DL
\end{DL}

\begin{gcotable}
property & C & p & q & - & count & - \\
\end{gcotable}

\section{Provenance}

Any provenance information must be available for instances of given classes.

\subsection{Simple Example}

For \emph{publications}, any provenance information must be available.

\begin{DL}
Not expressible in DL!
\end{DL}

\begin{gcotable}
class & Publication & - & - & - & provenance & - \\
\end{gcotable}

\section{Data Model Consistency}

Is the data consistent with the intended semantics of the data model?
Such validation rules ensure the integrity of the data according to the data model.

\begin{DL}
Not expressible in DL!
\end{DL}

\section{Structure}

SKOS is based on RDF, which is a graph-based data model. Therefore we can concentrate on the vocabulary's graph-based structure for assessing the quality of SKOS vocabularies and apply graph- and network-analysis techniques. 

\begin{DL}
Not expressible in DL!
\end{DL}

\section{Labeling and Documentation}

\begin{DL}
Not expressible in DL!
\end{DL}

\section{Vocabulary}

Vocabularies should not invent any new terms or use deprecated elements. 

\subsection{Simple Example}

\begin{DL}
Not expressible in DL!
\end{DL}

\begin{gcotable}
class & $\top$ & - & - & - & vocabulary & - \\
\end{gcotable}

\section{HTTP URI Scheme Violation}

\subsection{Simple Example}

\begin{DL}
Not expressible in DL!
\end{DL}

\begin{gcotable}
class & $\top$ & - & - & - & HTTP URI Scheme Violation & - \\
\end{gcotable}

\section{Evaluation}

\subsection{Evaluation of Constraint Languages}
\label{sec:evaluation-1}

We evaluated to which extend the most promising five constraint languages fulfill each requirement.
Tilde means that this constraint may be fulfilled by that particular constraint language - either by limitations, workarounds, or extensions.
We also evaluated if a specific constraint is fulfilled by OWL 2 QL or if the more expressive OWL 2 DL is needed. 
Inferencing may be performed prior to validating constraints. This is marked with an asterisk.

\begin{evaluation-overall}
*Subsumption & \ding{55} & $\checkmark$ & $\checkmark$ & $\sim$ & $\checkmark$ & $\checkmark$ \\
*Class Equivalence & \ding{55} & $\checkmark$ & $\checkmark$ & \ding{55} & \ding{55} & $\checkmark$ \\
*Sub Properties & \ding{55} & $\checkmark$ & $\checkmark$ & \ding{55} & \ding{55} & $\checkmark$ \\
*Property Domains & \ding{55} & $\checkmark$ & $\checkmark$ & \ding{55} & \ding{55} & $\checkmark$ \\
*Property Ranges & \ding{55} & $\checkmark$ & $\checkmark$ & \ding{55} & \ding{55} & $\checkmark$ \\
*Inverse Object Properties & \ding{55} & $\checkmark$ & $\checkmark$ & $\sim$ & \ding{55} & $\checkmark$ \\
*Symmetric Object Properties & \ding{55} & $\checkmark$ & $\checkmark$ & \ding{55} & \ding{55} & $\checkmark$ \\
*Asymmetric Object Properties & \ding{55} & $\checkmark$ & $\checkmark$ & \ding{55} & \ding{55} & $\checkmark$ \\
*Reflexive Object Properties & \ding{55} & $\checkmark$ & $\checkmark$ & \ding{55} & \ding{55} & $\checkmark$ \\
*Irreflexive Object Properties & \ding{55} & $\checkmark$ & $\checkmark$ & \ding{55} & \ding{55} & $\checkmark$ \\
Disjoint Properties & \ding{55} & $\checkmark$ & $\checkmark$ & \ding{55} & \ding{55} & $\checkmark$ \\
Disjoint Classes & \ding{55} & $\checkmark$ & $\checkmark$ & \ding{55} & \ding{55} & $\checkmark$ \\
Context-Sp. Property Groups & \ding{55} & $\sim$ & $\sim$ & $\checkmark$ & $\checkmark$ & $\checkmark$ \\
Context-Sp. Inclusive OR of P. & \ding{55} & $\sim$ & $\sim$ & \ding{55} & \ding{55} & $\checkmark$ \\
Context-Sp. Inclusive OR of P. Groups & \ding{55} & $\sim$ & $\sim$ & \ding{55} & \ding{55} & $\checkmark$ \\
Recursive Queries & $\checkmark$ & $\checkmark$ & $\checkmark$ & $\checkmark$ & $\checkmark$ & $\sim$ \\
Individual Inequality & \ding{55} & $\checkmark$ & $\checkmark$ & \ding{55} & \ding{55} & $\checkmark$ \\
*Equivalent Properties & \ding{55} & $\checkmark$ & $\checkmark$ & \ding{55} & \ding{55} & $\checkmark$ \\
Property Assertions & \ding{55} & $\checkmark$ & $\sim$ & \ding{55} & \ding{55} & $\checkmark$ \\
Data Property Facets & \ding{55} & $\checkmark$ & $\checkmark$ & \ding{55} & \ding{55} & $\checkmark$ \\
Literal Pattern Matching & \ding{55} & $\checkmark$ & \ding{55} & \ding{55} & \ding{55} & $\checkmark$ \\
Negative Literal Pattern Matching & \ding{55} & $\checkmark$ & \ding{55} & \ding{55} & \ding{55} & $\checkmark$ \\
*Object Property Paths & \ding{55} & $\checkmark$ & \ding{55} & \ding{55} & \ding{55} & $\checkmark$ \\
*Intersection & \ding{55} & $\checkmark$ & \ding{55} & $\checkmark$ & $\checkmark$ & $\checkmark$ \\
*Disjunction & \ding{55} & $\checkmark$ & \ding{55} & \ding{55} & \ding{55} & $\checkmark$ \\
*Negation & \ding{55} & $\checkmark$ & \ding{55} & \ding{55} & \ding{55} & $\checkmark$ \\
*Existential Quantifications & \ding{55} & $\checkmark$ & \ding{55} & $\sim$ & $\sim$ & $\checkmark$ \\
*Universal Quantifications & \ding{55} & $\checkmark$ & \ding{55} & \ding{55} & \ding{55} & $\checkmark$ \\
*Minimum Unqualified Cardinality & $\checkmark$ & $\checkmark$ & \ding{55} & $\sim$ & $\checkmark$ & $\checkmark$ \\
*Minimum Qualified Cardinality & $\checkmark$ & $\checkmark$ & \ding{55} & $\sim$ & $\checkmark$ & $\checkmark$ \\
*Maximum Unqualified Cardinality & $\checkmark$ & $\checkmark$ & \ding{55} & $\sim$ & $\checkmark$ & $\checkmark$ \\
*Maximum Qualified Cardinality & $\checkmark$ & $\checkmark$ & \ding{55} & $\sim$ & $\checkmark$ & $\checkmark$ \\
*Exact Unqualified Cardinality & $\checkmark$ & $\checkmark$ & \ding{55} & $\sim$ & $\checkmark$ & $\checkmark$ \\
*Exact Qualified Cardinality & $\checkmark$ & $\checkmark$ & \ding{55} & $\sim$ & $\checkmark$ & $\checkmark$ \\
*Transitive Object Properties & \ding{55} & $\checkmark$ & \ding{55} & \ding{55} & \ding{55} & $\checkmark$ \\
Context-Sp. Exclusive OR of P. & \ding{55} & $\checkmark$ & \ding{55} & \ding{55} & $\checkmark$ & $\checkmark$ \\
Context-Sp. Exclusive OR of P. Groups & \ding{55} & $\sim$ & \ding{55} & $\checkmark$ & $\checkmark$ & $\checkmark$ \\
Allowed Values & $\checkmark$ & $\checkmark$ & \ding{55} & $\checkmark$ & $\checkmark$& $\checkmark$ \\
Not Allowed Values & \ding{55} & $\checkmark$ & \ding{55} & \ding{55} & $\checkmark$ & $\checkmark$ \\
Literal Ranges & \ding{55} & $\checkmark$ & \ding{55} & \ding{55} & \ding{55} & $\checkmark$ \\
Negative Literal Ranges & \ding{55} & $\checkmark$ & \ding{55} & \ding{55} & \ding{55} & $\checkmark$ \\
Required Properties & $\checkmark$ & $\checkmark$ & \ding{55} & $\checkmark$ & $\checkmark$ & $\checkmark$ \\
Optional Properties & $\checkmark$ & $\checkmark$ & \ding{55} & $\checkmark$ & $\checkmark$ & $\checkmark$ \\
Repeatable Properties & $\checkmark$ & $\checkmark$ & \ding{55} & $\checkmark$ & $\checkmark$ & $\checkmark$ \\
Negative Property Constraints & \ding{55} & $\checkmark$ & \ding{55} & \ding{55} & $\checkmark$ & $\checkmark$ \\
*Individual Equality & \ding{55} & $\checkmark$ & \ding{55} & \ding{55} & \ding{55} & $\checkmark$ \\
*Functional Properties & \ding{55} & $\checkmark$ & \ding{55} & \ding{55} & \ding{55} & $\checkmark$ \\
*Inverse-Functional Properties & \ding{55} & $\checkmark$ & \ding{55} & \ding{55} & \ding{55} & $\checkmark$ \\
*Value Restrictions & $\checkmark$ & $\checkmark$ & \ding{55} & $\checkmark$ & $\checkmark$ & $\checkmark$ \\
*Self Restrictions & \ding{55} & $\checkmark$ & \ding{55} & \ding{55} & \ding{55} & $\checkmark$ \\
*Primary Key Properties & \ding{55} & $\checkmark$ & \ding{55} & \ding{55} & \ding{55} & $\checkmark$ \\
*Class-Specific Property Range & $\checkmark$ & $\checkmark$ & \ding{55} & $\checkmark$ & $\checkmark$ & $\checkmark$ \\
*Class-Sp. Reflexive Object P. & \ding{55} & $\checkmark$ & \ding{55} & \ding{55} & \ding{55} & $\checkmark$ \\
Membership in Controlled Vocabularies & $\checkmark$ & \ding{55} & \ding{55} & \ding{55} & \ding{55} & $\checkmark$ \\
IRI Pattern Matching & \ding{55} & \ding{55} & \ding{55} & \ding{55} & $\checkmark$ & $\checkmark$ \\
Literal Value Comparison & \ding{55} & \ding{55} & \ding{55} & \ding{55} & $\checkmark$ & $\checkmark$ \\
Ordering & \ding{55} & \ding{55} & \ding{55} & \ding{55} & \ding{55} & $\checkmark$ \\
Validation Levels & \ding{55} & \ding{55} & \ding{55} & \ding{55} & \ding{55} & $\checkmark$ \\
String Operations & \ding{55} & \ding{55} & \ding{55} & \ding{55} & \ding{55} & $\checkmark$ \\
\end{evaluation-overall}

\begin{evaluation-overall}
Context-Specific Valid Classes & \ding{55} & \ding{55} & \ding{55} & \ding{55} & \ding{55} & $\checkmark$ \\
Context-Specific Valid Properties & \ding{55} & \ding{55} & \ding{55} & \ding{55} & \ding{55} & $\checkmark$ \\
*Default Values & \ding{55} & \ding{55} & \ding{55} & $\checkmark$ & \ding{55} & $\checkmark$ \\
Mathematical Operations & \ding{55} & \ding{55} & \ding{55} & \ding{55} & \ding{55} & $\checkmark$ \\
Language Tag Matching & \ding{55} & \ding{55} & \ding{55} & \ding{55} & \ding{55} & $\checkmark$ \\
Language Tag Cardinality & \ding{55} & \ding{55} & \ding{55} & \ding{55} & \ding{55} & $\checkmark$ \\
Whitespace Handling & \ding{55} & \ding{55} & \ding{55} & \ding{55} & \ding{55} & $\checkmark$ \\
HTML Handling & \ding{55} & \ding{55} & \ding{55} & \ding{55} & \ding{55} & $\checkmark$ \\
Conditional Properties & \ding{55} & \ding{55} & \ding{55} & \ding{55} & \ding{55} & $\checkmark$ \\
Recommended Properties & \ding{55} & \ding{55} & \ding{55} & \ding{55} & \ding{55} & $\checkmark$ \\
Handle RDF Collections & \ding{55} & \ding{55} & \ding{55} & \ding{55} & \ding{55} & $\checkmark$ \\
Value is Valid for Datatype & \ding{55} & \ding{55} & \ding{55} & \ding{55} & \ding{55} & $\checkmark$ \\
Use Sub-Super Relations in Validation & \ding{55} & \ding{55} & \ding{55} & \ding{55} & \ding{55} & $\checkmark$ \\
*Cardinality Shortcuts & \ding{55} & $\checkmark$ & \ding{55} & $\checkmark$ & $\checkmark$ & $\checkmark$ \\
Aggregations & \ding{55} & \ding{55} & \ding{55} & \ding{55} & \ding{55} & $\checkmark$ \\

*Class-Specific Irreflexive Object Properties & \ding{55} & $\checkmark$ & \ding{55} & \ding{55} & \ding{55} & $\checkmark$ \\
Provenance & \ding{55} & \ding{55} & \ding{55} & \ding{55} & \ding{55} & $\checkmark$ \\
Data Model Consistency & \ding{55} & \ding{55} & \ding{55} & \ding{55} & \ding{55} & $\checkmark$ \\
Structure & \ding{55} & \ding{55} & \ding{55} & \ding{55} & \ding{55} & $\checkmark$ \\  
Labeling and Documentation & \ding{55} & \ding{55} & \ding{55} & \ding{55} & \ding{55} & $\checkmark$ \\
Vocabulary & \ding{55} & \ding{55} & \ding{55} & \ding{55} & \ding{55} & $\checkmark$ \\
HTTP URI Scheme Violation & \ding{55} & \ding{55} & \ding{55} & \ding{55} & \ding{55} & $\checkmark$ \\
\end{evaluation-overall}

\subsection{Evaluation of Constraints Classes}
\label{sec:evaluation-2}

We evaluated to which extend the most promising five constraint languages fulfill each of the overall 74 requirements to formulate RDF constraints \cite{BoschNolleAcarEckert2015}.
If a constraint can be expressed in DL, we added the mapping to DL and to the generic constraint.
If a constraint cannot be expressed in DL, we only added the mapping to the generic constraint.
Therefore, we show that each constraint can be mapped to a generic constraint.
The following table shows the absolute numbers and the relative percentages for each of the three dimensions to classify constraints:

\begin{evaluation-generic-overview}
Property Constraints & 49 & 60 (60.49) \\
Class Constraints & 20 & 25 (24.96) \\
Property and Class Constraints & 12 & 15 (14.81) \\
\hline
Simple Constraints & 49 & 60 (60.49) \\
Simple Constraints (Syntactic Sugar) & 11 & 14 (13.58) \\
Complex Constraints & 21 & 26 (25.93) \\
\hline
DL Expressible & 52 & 64 (64.2) \\
DL Not Expressible & 29 & 36 (35.80) \\
\hline
Total & 81 & 100 \\
\end{evaluation-generic-overview}

legend of the detailed evaluation:
\begin{itemize}
	\item \ms{property constraint:} property constraint ($\checkmark$) vs. class constraint (\ding{55}) vs. property and class constraints ($\sim$) 
	\item \ms{simple constraint:} simple constraint ($\checkmark$) vs. simple constraint [syntactic sugar] ($\sim$) vs. complex constraint (\ding{55})
	\item \ms{DL:} expressible in DL ($\checkmark$) vs. not expressible in DL (\ding{55})
\end{itemize}

\begin{evaluation-generic}
*Subsumption & \ding{55} & $\checkmark$ & $\checkmark$ \\ 
*Class Equivalence & \ding{55} & \ding{55} & $\checkmark$ \\ 
*Sub Properties & $\checkmark$ & $\checkmark$ & $\checkmark$ \\  
*Property Domains & $\checkmark$ & $\sim$ & $\checkmark$ \\ 
*Property Ranges & $\checkmark$ & $\sim$ & $\checkmark$ \\ 
*Inverse Object Properties & $\checkmark$ & $\checkmark$ & $\checkmark$ \\  
*Symmetric Object Properties & $\checkmark$ & $\checkmark$ & $\checkmark$ \\  
*Asymmetric Object Properties & $\checkmark$ & $\sim$ & $\checkmark$ \\ 
*Reflexive Object Properties & $\checkmark$ & $\sim$ & $\checkmark$ \\ 
*Irreflexive Object Properties & $\checkmark$ & $\sim$ & $\checkmark$ \\
Disjoint Properties & $\checkmark$ & $\checkmark$ & $\checkmark$ \\  
Disjoint Classes & \ding{55} & \ding{55} & $\checkmark$ \\ 
Context-Sp. Property Groups & $\sim$ & \ding{55} & $\checkmark$ \\  
Context-Sp. Inclusive OR of P. & $\sim$ & \ding{55} & $\checkmark$ \\
Context-Sp. Inclusive OR of P. Groups & $\sim$ & \ding{55} & $\checkmark$ \\ 
Recursive Queries & $\checkmark$ & $\checkmark$ & $\checkmark$ \\  
Individual Inequality & \ding{55} & \ding{55} & $\checkmark$ \\  
*Equivalent Properties & $\checkmark$ & \ding{55} & $\checkmark$ \\   
Property Assertions & $\checkmark$ & $\checkmark$ & $\checkmark$ \\  
Data Property Facets & \ding{55} & $\checkmark$ & \ding{55} \\  
Literal Pattern Matching & \ding{55} & $\checkmark$ & \ding{55} \\   
Negative Literal Pattern Matching & \ding{55} & \ding{55} & \ding{55} \\ 
*Object Property Paths & $\checkmark$ & $\checkmark$ & $\checkmark$ \\  
*Intersection & \ding{55} & $\checkmark$ & $\checkmark$ \\  
*Disjunction & \ding{55} & $\checkmark$ & $\checkmark$ \\ 
*Negation & \ding{55} & $\checkmark$ & $\checkmark$ \\ 
*Existential Quantifications & $\checkmark$ & $\checkmark$ & $\checkmark$ \\  
*Universal Quantifications & $\checkmark$ & $\checkmark$ & $\checkmark$ \\  
*Minimum Unqualified Cardinality & $\checkmark$ & $\checkmark$ & $\checkmark$ \\  
*Minimum Qualified Cardinality & $\checkmark$ & $\checkmark$ & $\checkmark$ \\  
*Maximum Unqualified Cardinality & $\checkmark$ & $\checkmark$ & $\checkmark$ \\  
*Maximum Qualified Cardinality & $\checkmark$ & $\checkmark$ & $\checkmark$ \\  
*Exact Unqualified Cardinality & $\checkmark$ & $\sim$ & $\checkmark$ \\ 
*Exact Qualified Cardinality & $\checkmark$ & $\sim$ & $\checkmark$ \\ 
*Transitive Object Properties & $\checkmark$ & $\checkmark$ & $\checkmark$ \\  
Context-Sp. Exclusive OR of P. & $\sim$ & \ding{55} & $\checkmark$ \\ 
Context-Sp. Exclusive OR of P. Groups & $\sim$ & \ding{55} & $\checkmark$ \\
Allowed Values & \ding{55} & $\checkmark$ & $\checkmark$ \\  
Not Allowed Values & \ding{55} & \ding{55} & $\checkmark$ \\  
Literal Ranges & \ding{55} & \ding{55} & \ding{55} \\ 
Negative Literal Ranges & \ding{55} & \ding{55} & \ding{55} \\ 
Required Properties & $\checkmark$ & $\checkmark$ & $\checkmark$ \\  
Optional Properties & $\checkmark$ & $\sim$ & $\checkmark$ \\
Repeatable Properties & $\checkmark$ & $\checkmark$ & $\checkmark$ \\  
Negative Property Constraints & $\sim$ & \ding{55} & $\checkmark$ \\ 
*Individual Equality & \ding{55} & \ding{55} & $\checkmark$ \\  
*Functional Properties & $\checkmark$ & $\checkmark$ & $\checkmark$ \\ 
*Inverse-Functional Properties & $\checkmark$ & \ding{55} & $\checkmark$ \\ 
*Value Restrictions & $\checkmark$ & $\checkmark$ & $\checkmark$ \\ 
*Self Restrictions & $\sim$ & \ding{55} & $\checkmark$ \\
*Primary Key Properties & $\checkmark$ & $\sim$ & $\checkmark$ \\ 
*Class-Specific Property Range & $\checkmark$ & $\sim$ & $\checkmark$ \\ 
*Class-Sp. Reflexive Object P. & $\checkmark$ & $\checkmark$ & $\checkmark$ \\  
Membership in Controlled Vocabularies & $\sim$ & \ding{55} & $\checkmark$ \\ 
IRI Pattern Matching & \ding{55} & $\checkmark$ & \ding{55} \\   
Literal Value Comparison & $\checkmark$ & $\checkmark$ & \ding{55} \\   
Ordering & $\checkmark$ & $\checkmark$ & \ding{55} \\ 
Validation Levels & $\sim$ & $\checkmark$ & \ding{55} \\
String Operations & $\checkmark$ & $\checkmark$ & \ding{55} \\ 
\end{evaluation-generic}

\begin{evaluation-generic}
Context-Specific Valid Classes & \ding{55} & $\checkmark$ & \ding{55} \\   
Context-Specific Valid Properties & $\checkmark$ & $\checkmark$ & \ding{55} \\ 
*Default Values & $\checkmark$ & $\checkmark$ & \ding{55} \\ 
Mathematical Operations & $\checkmark$ & $\checkmark$ & \ding{55} \\ 
Language Tag Matching & $\checkmark$ & $\checkmark$ & \ding{55} \\ 
Language Tag Cardinality & $\checkmark$ & $\checkmark$ & \ding{55} \\ 
Whitespace Handling & $\checkmark$ & $\checkmark$ & \ding{55} \\ 
HTML Handling & $\checkmark$ & $\checkmark$ & \ding{55} \\ 
Conditional Properties & $\checkmark$ & $\checkmark$ & $\checkmark$ \\  
Recommended Properties & $\checkmark$ & $\checkmark$ & \ding{55} \\ 
Handle RDF Collections & $\checkmark$ & $\checkmark$ & \ding{55} \\ 
Value is Valid for Datatype & $\checkmark$ & $\checkmark$ & \ding{55} \\ 
Use Sub-Super Relations in Validation & $\checkmark$ & $\checkmark$ & \ding{55} \\ 
*Cardinality Shortcuts & $\checkmark$ & $\checkmark$ & $\checkmark$ \\  
Aggregations & $\checkmark$ & $\checkmark$ & \ding{55} \\ 

*Class-Specific Irreflexive Object Properties & $\checkmark$ & $\sim$ & $\checkmark$ \\
Provenance & \ding{55} & $\checkmark$ & \ding{55} \\
Data Model Consistency & $\sim$ & \ding{55} & \ding{55} \\
Structure & $\sim$ & \ding{55} & \ding{55} \\ 
Labeling and Documentation & $\sim$ & \ding{55} & \ding{55} \\
Vocabulary & \ding{55} & $\checkmark$ & \ding{55} \\
HTTP URI Scheme Violation & \ding{55} & $\checkmark$ & \ding{55} \\
\end{evaluation-generic}

Constraints can be classified as \ms{property constraints} and \ms{class constraints}.
Two thirds of the total amount of constraints are property constraints, one fifth are class constraints, and approx. 10\% are composed of both property and class constraints.
Constraints may be either atomic (\ms{simple constraints}) or created out of simple and/or complex constraints (\ms{complex constraints}).
Almost two thirds are simple constraints, a quarter are complex constraints.
Almost 15 percent are complex constraints which can be formulated as simple constraints when using them in terms of syntactic sugar.
Constraints can either be expressible in DL or not.
The majority - nearly 70\% - of the overall constraints are expressible in DL.

\subsection{CWA and UNA Dependency}

\begin{itemize}
	\item Dependent on the \emph{CWA}: do further triples change the validation result?
	\item Dependent on the \emph{UNA}: Do validation results changes in case two resources are the same?
\end{itemize}

\begin{table}[H]
    \begin{center}
    \begin{tabular}{@{}lcc@{}}
           & \multicolumn{2}{c}{\textbf{Dependency}}
    \\  \cmidrule{2-3}
    \\       \textbf{Constraint Types}
           & \rot{\emph{CWA}}
           & \rot{\emph{UNA}}
    \\ \midrule
		*Subsumption & $\checkmark$ & $\checkmark$ \\
		*Class Equivalence & $\checkmark$ & $\checkmark$ \\
		*Sub Properties & $\checkmark$  & $\checkmark$ \\
		*Property Domains & $\checkmark$ & $\checkmark$ \\
		*Property Ranges & $\checkmark$ & $\checkmark$ \\
		*Inverse Object Properties & $\checkmark$ & $\checkmark$ \\
		*Symmetric Object Properties & $\checkmark$ & $\checkmark$ \\
		*Asymmetric Object Properties & \ding{55} & \ding{55} \\
		*Reflexive Object Properties & $\checkmark$ & $\checkmark$ \\
		*Irreflexive Object Properties & \ding{55} & \ding{55} \\
		Disjoint Properties & \ding{55} & \ding{55} \\
		Disjoint Classes & \ding{55} & $\checkmark$ \\
		Context-Sp. Property Groups & $\checkmark$ & $\checkmark$ \\
		Context-Sp. Inclusive OR of P. & $\checkmark$ & $\checkmark$ \\
		Context-Sp. Inclusive OR of P. Groups & $\checkmark$ & $\checkmark$ \\
		Recursive Queries & \ding{55} & \ding{55} \\
		Individual Inequality & \ding{55} & \ding{55} \\
		*Equivalent Properties & $\checkmark$ & $\checkmark$ \\
		Property Assertions & $\checkmark$ & $\checkmark$ \\
		Data Property Facets & \ding{55} & \ding{55} \\
		Literal Pattern Matching & \ding{55} & \ding{55} \\
		Negative Literal Pattern Matching & \ding{55} & \ding{55} \\
		*Object Property Paths & $\checkmark$ & $\checkmark$ \\
		*Intersection & $\checkmark$ & $\checkmark$ \\
		*Disjunction & $\checkmark$ & $\checkmark$ \\
		*Negation & \ding{55} & \ding{55} \\
		*Existential Quantifications & $\checkmark$ & $\checkmark$ \\
		*Universal Quantifications & \ding{55} & $\checkmark$ \\
		*Minimum Unqualified Cardinality & $\checkmark$ & $\checkmark$ \\
		*Minimum Qualified Cardinality & $\checkmark$ & $\checkmark$ \\
		*Maximum Unqualified Cardinality & \ding{55} & $\checkmark$ \\
		*Maximum Qualified Cardinality & \ding{55} & $\checkmark$ \\
		*Exact Unqualified Cardinality & $\checkmark$ & $\checkmark$ \\
		*Exact Qualified Cardinality & $\checkmark$ & $\checkmark$ \\
		*Transitive Object Properties & $\checkmark$ & $\checkmark$ \\
		Context-Sp. Exclusive OR of P. & \ding{55} & $\checkmark$ \\
		Context-Sp. Exclusive OR of P. Groups & \ding{55} & $\checkmark$ \\	
		Allowed Values & \ding{55} & $\checkmark$ \\
		Not Allowed Values & \ding{55} & $\checkmark$ \\
		Literal Ranges & \ding{55} & \ding{55} \\
		Negative Literal Ranges & \ding{55} & \ding{55} \\
		Required Properties & $\checkmark$ & $\checkmark$ \\
		Optional Properties & \ding{55} & \ding{55} \\
		Repeatable Properties & $\checkmark$ & \ding{55} \\
		Negative Property Constraints & \ding{55} & $\checkmark$ \\
		*Individual Equality & $\checkmark$ & \ding{55} \\
		*Functional Properties & $\checkmark$ & $\checkmark$ \\
		*Inverse-Functional Properties & $\checkmark$ & $\checkmark$ \\
		*Value Restrictions & $\checkmark$ & $\checkmark$ \\
		*Self Restrictions & $\checkmark$ & $\checkmark$ \\
    \bottomrule
    \end{tabular}
		\label{tab:evaluation-cwa-una-dependency-1}
    \end{center}
\end{table}

\begin{table}[H]
    \begin{center}
    \begin{tabular}{@{}lcc@{}}
           & \multicolumn{2}{c}{\textbf{Dependency}}
    \\  \cmidrule{2-3}
    \\       \textbf{Constraint Types}
           & \rot{\emph{CWA}}
           & \rot{\emph{UNA}}
    \\ \midrule
		*Primary Key Properties & $\checkmark$ & $\checkmark$ \\
		*Class-Specific Property Range & $\checkmark$ & $\checkmark$ \\
		*Class-Sp. Reflexive Object P. & $\checkmark$ & $\checkmark$ \\
		Membership in Controlled Vocabularies & $\checkmark$ & $\checkmark$ \\
		IRI Pattern Matching & \ding{55} & $\checkmark$ \\
		Literal Value Comparison & \ding{55} & \ding{55} \\
		Ordering & $\checkmark$ & \ding{55} \\
		Validation Levels & $\checkmark$ & \ding{55} \\
		String Operations & \ding{55} & \ding{55} \\
		Context-Specific Valid Classes & \ding{55} & \ding{55} \\
		Context-Specific Valid Properties & \ding{55} & \ding{55} \\
		*Default Values & $\checkmark$ & $\checkmark$ \\
		Mathematical Operations & \ding{55} & \ding{55} \\
		Language Tag Matching & $\checkmark$ & \ding{55} \\
		Language Tag Cardinality & $\checkmark$ & $\checkmark$ \\
		Whitespace Handling & \ding{55} & \ding{55} \\
		HTML Handling & \ding{55} & \ding{55} \\
		Conditional Properties & $\checkmark$ & $\checkmark$ \\
		Recommended Properties & $\checkmark$ & $\checkmark$ \\
		Handle RDF Collections & \ding{55} & \ding{55} \\
		Value is Valid for Datatype & \ding{55} & \ding{55} \\
		Use Sub-Super Relations in Validation & \ding{55} & $\checkmark$ \\
		*Cardinality Shortcuts & $\checkmark$ & $\checkmark$ \\ 
		Aggregations & \ding{55} & \ding{55} \\
		*Class-Specific Irreflexive Object Properties & \ding{55} & $\checkmark$ \\
		Provenance & $\checkmark$ & $\checkmark$ \\ 
		Data Model Consistency & $\checkmark$ & $\checkmark$ \\ 
		Structure & $\checkmark$ & $\checkmark$ \\  
		Labeling and Documentation & $\checkmark$ & $\checkmark$ \\ 
		Vocabulary & $\checkmark$ & $\checkmark$ \\
		HTTP URI Scheme Violation & \ding{55} & $\checkmark$ \\
    \bottomrule
    \end{tabular}
		\label{tab:evaluation-cwa-una-dependency-2}
    \end{center}
\end{table}

\subsection{Constraining Elements of Constraint Types}

\begin{itemize}
  \item Constraints of 64\% of all constraint type can be expressed using logical constructs from description logics.	
	\item For a constraint types, there can be multiple constraining elements
	\item The validation of each constraining element has to be implemented
	\item We use SPARQL as extension mechanism to express constraints of 3 constraint types which cannot be expressed otherwise as they are too complex to represent them using a high-level constraint language
\end{itemize}

\begin{table}[H]
    \begin{center}
    \begin{tabular}{lll}
           \multicolumn{1}{l}{\textbf{Constraint Type}} & \multicolumn{1}{l}{\textbf{DL}} & \multicolumn{1}{l}{\textbf{Constraining Elements}}			 
    \\ \midrule
		
*Subsumption & $\sqsubseteq$ & sub-class \\
*Class Equivalence & $\equiv$ & equivalent class \\
*Sub Properties & $\sqsubseteq$ & sub-property \\
*Property Domains & $\exists$, $\sqsubseteq$ & property domain \\
*Property Ranges & $\sqsubseteq$, $\forall$ & property range \\
*Inverse Object Properties & inverse & inverse property \\
*Symmetric Object Properties & symmetric & symmetric property \\
*Asymmetric Object Properties & asymmetric & asymmetric property \\
*Reflexive Object Properties & reflexive & reflexive property \\
*Irreflexive Object Properties & reflexive, $\neg$ & irreflexive property \\
Disjoint Properties & $\sqsubseteq$, $\neg$ & disjoint property \\
Disjoint Classes & $\sqcap$, $\sqsubseteq$ & disjoint class \\
Context-Sp. Property Groups & $\sqcap$ & property group \\
Context-Sp. Inclusive OR of P. & $\sqcup$ & inclusive or \\
Context-Sp. Inclusive OR of P. Groups & $\sqcup$, $\sqcap$ & inclusive or \\
Recursive Queries & - & recursive \\
Individual Inequality & $\ne$ & different individuals \\
*Equivalent Properties & $\equiv$ & equivalent properties \\
Property Assertions & = or $\ne$ & =, $\ne$ \\
Data Property Facets & - & SPARQL functions: \\
                        && REGEX, STRLEN \\
												&& XML Schema constraining facets: \\
                        && xsd:length, xsd:minLength, xsd:maxLength, xsd:pattern, \\
												&& xsd:enumeration, xsd:whiteSpace, xsd:maxInclusive, \\
												&& xsd:maxExclusive,xsd:minExclusive, xsd:minInclusive, \\
												&& xsd:totalDigits, xsd:fractionDigits \\
Literal Pattern Matching & - & REGEX, xsd:pattern \\
Negative Literal Pattern Matching & - & REGEX, xsd:pattern and $\neg$ \\ 
*Object Property Paths & $\sqsubseteq$ & property path \\
*Intersection & $\sqcap$ & intersection \\
*Disjunction & $\sqcup$ & disjunction \\
*Negation & $\neg$ & negation \\
*Existential Quantification & $\exists$ & existential quantification \\
*Universal Quantification & $\forall$ & universal quantification \\
*Minimum Unqualified Cardinality & $\geq$ &  minimum unqualified cardinality restriction \\
*Minimum Qualified Cardinality & $\geq$ &  minimum qualified cardinality restriction \\
*Maximum Unqualified Cardinality & $\leq$ &  maximum unqualified cardinality restriction \\
*Maximum Qualified Cardinality & $\leq$ &  maximum qualified cardinality restriction \\
*Exact Unqualified Cardinality & ( $\geq$, $\leq$, $\sqcap$ ) or = &  exact unqualified cardinality restriction \\
*Exact Qualified Cardinality & ( $\geq$, $\leq$, $\sqcap$ ) or = &  exact qualified cardinality restriction \\
*Transitive Object Properties & $\sqsubseteq$ & transitive property \\
Context-Sp. Exclusive OR of P. & $\neg$, $\sqcap$, $\sqcup$ & exclusive or \\
Context-Sp. Exclusive OR of P. Groups & $\neg$, $\sqcap$, $\sqcup$ & exclusive or \\
Allowed Values & $\sqcup$ & allowed values \\
Not Allowed Values & $\sqcup$, $\neg$ & not allowed values \\
Literal Ranges & - & xsd:minInclusive, xsd:maxExclusive \\
               & & xsd:maxInclusive, xsd:minExclusive \\
Negative Literal Ranges & - & $\neg$ and \\  
                        && xsd:minInclusive, xsd:maxExclusive \\
                        && xsd:maxInclusive, xsd:minExclusive \\
Required Properties & $\exists$ & required property \\	
	
    \bottomrule
    \end{tabular}
    \end{center}
\end{table}

\begin{table}[H]
    \begin{center}
    \begin{tabular}{lll}
           \multicolumn{1}{l}{\textbf{Constraint Type}} & \multicolumn{1}{l}{\textbf{DL}} & \multicolumn{1}{l}{\textbf{Constraining Elements}}			 
    \\ \midrule
		
Optional Properties & $\exists$, $\sqsubseteq$ & optional property \\
Repeatable Properties & $\geq$ & repeatable property \\
Negative Property Constraints & $\neg$, $\exists$ & negative properties \\
*Individual Equality & = & equal individuals \\
*Functional Properties & functional & functional property \\ 
*Inverse-Functional Properties & inverse, functional & inverse functional property \\
*Value Restrictions & $\exists$ & value restriction \\
*Self Restrictions & $\exists$ & self restriction \\
Primary Key Properties & inverse, functional & primary key \\
*Class-Specific Property Range & $\sqsubseteq$, $\exists$, $\neg$ & property range \\
*Class-Sp. Reflexive Object P. & reflexive & reflexive property \\
Membership in Controlled Vocabularies & $\forall$, $\sqcap$, $\sqcup$ & membership in controlled vocabularies \\
IRI Pattern Matching & - & IRI pattern matching \\
Literal Value Comparison & $>$ or $\geq$ or $<$ or & $>$, $\geq$, $<$, $\leq$, =, $\ne$ \\
                         & $\leq$ or = or $\ne$ & \\
Ordering & - & ordered properties, ordered values \\ 
Validation Levels & - & validation level \\ 
String Operations & - & SPARQL string functions: \\ 
                     && STRLEN, SUBSTR, UCASE, LCASE, \\
										 && STRSTARTS, STRENDS, CONTAINS, \\ 
										 && STRBEFORE, STRAFTER, ENCODE\_FOR\_URI, \\
										 && CONCAT, langMatches, REGEX, REPLACE \\
Context-Specific Valid Classes & - & context-specific valid classes \\
Context-Specific Valid Properties & - & context-specific valid properties \\
Default Values & - & default values \\
Mathematical Operations & - & addition, subtraction, multiplication, division \\
Language Tag Matching & - & language tag matching \\
Language Tag Cardinality & - & language tag minimum cardinality, \\
                            && language tag maximum cardinality, \\
													  && language tag exact cardinality \\
Whitespace Handling & - & no whitespaces \\
HTML Handling & - & no html \\
Conditional Properties & $\sqsubseteq$ & conditional property \\
Recommended Properties & - & recommended property \\
Handle RDF Collections & - & actions on RDF collections: \\
                          && append list element \\
Value is Valid for Datatype & - & value is valid for datatype \\
Use Sub-Super Relations in Validation & - & non redundant properties \\
*Cardinality Shortcuts & & optional \& non-repeatable property, \\
                       & & optional \& repeatable property, \\
											 & & mandatory \& non-repeatable property, \\
											 & & mandatory \& repeatable property \\
Aggregations & - & aggregations: \\
                 && count \\
*Class-Specific Irreflexive Object Properties & reflexive, $\neg$ & irreflexive property \\
Provenance & - & provenance information required \\   
Data Model Consistency & - & $<$SPARQL query$>$ \\    
Structure & - & $<$SPARQL query$>$ \\  
Labeling and Documentation & - & $<$SPARQL query$>$ \\   
Vocabulary & - & vocabulary \\  
HTTP URI Scheme Violation & - & HTTP URI scheme violation \\   
		
    \bottomrule
    \end{tabular}
    \end{center}
\end{table}

\section{Conclusion and Future Work}
There is no standard way to validate RDF data conforming to RDF constraints like XML Schemas serve to validate XML documents.
Two working groups currently try to achieve a solution for RDF validation - the W3C RDF Data Shapes working group and the DCMI RDF Application Profiles working group. 
We initiated a comprehensive database on RDF validation requirements: \url{http://purl.org/net/rdf-validation}.
The intention of this database is to collaboratively work on case studies, use cases, requirements, and solutions on RDF validation.
In this paper, we evaluated to which extend the five most promising constraint languages on being the standard (DSP, OWL2, ReSh, ShEx, and SPIN) fulfill each of the requirements to formulate RDF constraints.
Each of these requirements corresponds to a type of constraint.
The majority of the constraints can be expressed in DL, which serves as a logical underpinning of related requirements.
We developed an ontology to express any constraint generically, so that constrains expressed by a constraint language $\alpha$ can be transformed into constraints expressed by a constraint language $\beta$ without any information loss.
By expressing any constraint generically, we can provide a validation of the generically expressed constraint.
When specific constraints are then transformed into generic constraints, we can provide the validation of the semantically equivalent specific constraints (expressed by multiple constraint languages) out-of-the-box without any additional effort and without any difference in validation results. 
As not every constrains can be represented in DL, we need to represent \emph{constraints expressible by DL} as well as \emph{constraints not expressible by DL} by means of this ontology.
We shown in terms of an evaluation that any constraint can be expressed using the developed ontology.
As part of future work, we will continuously add, modify, and maintain case studies, use cases, requirements, and solutions within the RDF validation requirements database.

\bibliography{../../literature/literature}{}
\bibliographystyle{plain}
\setcounter{tocdepth}{1}
\end{document}